\documentclass[aps,prx,twocolumn,notitlepage,superscriptaddress,showpacs,floatfix,amsmath]{revtex4-2}

\usepackage{tipa} 
\usepackage{bbding} 
\usepackage{amsfonts} 
\usepackage{amssymb}
\usepackage{graphicx}
\usepackage[colorlinks]{hyperref} 
\usepackage{braket}  

\begin{document}


\title{Two-level approximation of transmons in quantum quench experiments}

\author{H. S. Yan} 
\altaffiliation[]{These authors contributed equally to this work.}
\affiliation{Beijing National Laboratory for Condensed Matter Physics,
	Institute of Physics, Chinese Academy of Sciences, Beijing 100190, China}
\affiliation{School of Physical Sciences, University of Chinese Academy of Sciences,
	Beijing 100190, China}

\author{Yong-Yi Wang}
\altaffiliation[]{These authors contributed equally to this work.}
\affiliation{Beijing National Laboratory for Condensed Matter Physics, Institute of
	Physics, Chinese Academy of Sciences, Beijing 100190, China} \affiliation{School of
	Physical Sciences, University of Chinese Academy of Sciences, Beijing 100190, China}

\author{S. K. Zhao} \affiliation{Beijing Academy of Quantum
	Information Sciences, Beijing 100193, China}

\author{Z. H. Yang}
\affiliation{Beijing National Laboratory for Condensed Matter Physics, Institute of
	Physics, Chinese Academy of Sciences, Beijing 100190, China} \affiliation{School of
	Physical Sciences, University of Chinese Academy of Sciences, Beijing 100190, China}

\author{Z. T. Wang}
\affiliation{Beijing National Laboratory for Condensed Matter Physics, Institute of
	Physics, Chinese Academy of Sciences, Beijing 100190, China} \affiliation{School of
	Physical Sciences, University of Chinese Academy of Sciences, Beijing 100190, China}

\author{Kai Xu}
\affiliation{Beijing National Laboratory for Condensed Matter Physics, Institute of
	Physics, Chinese Academy of Sciences, Beijing 100190, China} \affiliation{School of
	Physical Sciences, University of Chinese Academy of Sciences, Beijing 100190, China}

\author{Ye Tian}
\affiliation{Beijing National Laboratory for Condensed Matter Physics, Institute of
	Physics, Chinese Academy of Sciences, Beijing 100190, China} 

\author{H. F. Yu} \affiliation{Beijing Academy of Quantum
	Information Sciences, Beijing 100193, China}

\author{Heng Fan} \affiliation{Beijing National Laboratory
	for Condensed Matter Physics, Institute of Physics, Chinese Academy of Sciences, Beijing
	100190, China} \affiliation{School of Physical Sciences, University of Chinese Academy of
	Sciences, Beijing 100190, China} \affiliation{CAS Center for Excellence in Topological
	Quantum Computation, UCAS, Beijing 100190, China} \affiliation{Songshan Lake Materials
	Laboratory, Dongguan 523808, China}

\author{S. P. Zhao} \affiliation{Beijing National Laboratory
	for Condensed Matter Physics, Institute of Physics, Chinese Academy of Sciences, Beijing
	100190, China} \affiliation{School of Physical Sciences, University of Chinese Academy of
	Sciences, Beijing 100190, China} \affiliation{CAS Center for Excellence in Topological
	Quantum Computation, UCAS, Beijing 100190, China} \affiliation{Songshan Lake Materials
	Laboratory, Dongguan 523808, China}


\begin{abstract}
	
Quantum quench is a typical protocol in the study of nonequilibrium dynamics of
quantum many-body systems. Recently, a number of experiments with
superconducting transmon qubits are reported, in which the spin and hard-core
boson models with two energy levels on individual sites are used. The transmons
are a multilevel system and the coupled qubits are governed by the Bose-Hubbard
model. How well they can be approximated by a two-level system has been
discussed and analysed in different ways for specific experiments in the
literature. Here, we numerically investigate the accuracy and validity of the
two-level approximation for the multilevel transmons based on the concept of
Loschmidt echo. Using this method, we are able to calculate the fidelity decay
(i.e., the time-dependent overlap of evolving wave functions) due to the state
leakage to transmon high energy levels. We present the results for different
system Hamiltonians with various initial states, qubit coupling strength, and
external driving, and for two kinds of quantum quench experiments with time
reversal and time evolution in one direction. We show quantitatively the extent
to which the fidelity decays with time for changing coupling strength (or
on-site interaction over coupling strength) and filled particle number or
locations in the initial states under specific system Hamiltonians, which may
serve as a way for assessing the two-level approximation of transmons. Finally,
we compare our results with the reported experiments using transmon qubits.

\vspace{3mm}

\end{abstract}


\maketitle

\section{Introduction}

Quantum simulators and computers promise to fulfil tasks that are
computationally difficult or impossible for their classical
counterparts~\cite{Geo14, Alt21}. While scalable fault-tolerant digital quantum
devices will be developed in the future for universal applications, analog (or
hybrid analog-digital) devices are expected to play an important role for
special-purpose applications in the coming years and decades~\cite{Dal22}. They
have been widely used in the studies of condensed-matter physics and phase
transition, high-energy physics and cosmology, and nonequilibrium quantum
many-body dynamics in isolated systems~\cite{Geo14, Alt21}. For example, the
quantum quench dynamics (QQD), essential in the study of nonequilibrium
properties of isolated quantum systems~\cite{Pol11}, can be explored
by preparing an initial state, which evolves under a Hamiltonian with
instantaneously changed parameters and is finally measured. QQD is a kind of
problem most challenging for classical computers and may serve as a good example
for the demonstration of practical quantum advantage~\cite{Dal22}.

Superconducting quantum processors based on transmon qubits~\cite{Koch07} are an
excellent platform for the quantum simulation of many-body physics~\cite{Xu2018,
	Roushan2017, Ma2019, Yan2019, Xu2020, Gong2021, Chen2021, Yanay20, Braum2021,
	Kara2022, Zhao21, Zhao22, Zhu2022} and demonstration of quantum
advantage~\cite{Arute2019, Wu21}. The transmon qubits have nonequidistant energy
levels, among which the two lowest levels form the computational
subspace~\cite{Krantz19}. In many applications, the initial states are prepared
in the two-level subspace. However, high qubit levels, or the noncomputational
states are known to be inevitably involved, and the term `state leakage' is
often used for their unwanted populations (i.e., the probability of finding
qubit in the high level states). As will be seen in this work, they can be
populated quantum mechanically at the very beginning of a quantum quench process
with the population depending on the initial states and other experimental
parameters such as the qubit coupling strength and applied driving. Some
fundamental operations like the single-qubit gates, entangling gates, and
measurements are also known to populate the qubit noncomputational levels
\cite{Mce21}.

The many-body spin and hard-core boson models with two energy levels on
individual sites have a long history, whose interesting physics has been a
subject of continuous investigation~\cite{Geo14,Suz13,Mon21,Caz11}. In recent
years, they have been frequently used in the QQD experiments with
superconducting circuits~\cite{Xu2018, Yan2019, Xu2020, Gong2021, Chen2021,
	Yanay20, Braum2021, Kara2022, Zhao21, Zhao22}. Due to the multilevel nature of
superconducting qubits, it is important to estimate the influence of the qubit
high levels in the experiment for the description in terms of these models. In
the previous works, this has been studied by evaluating the particle excitations
in the qubit high levels~\cite{Yan2019}, by considering the Loschmidt echo (LE)
and examining the deviations of the quantities under study from their ideal
values~\cite{Braum2021}, and by the satisfactoriness of the theoretical fits to
the experimental data~\cite{Zhao21}.

In this work,  by numerically simulating the QQD experiments of a typical
transmon-type qubit system, we study the influence of state leakage on quantum
state evolutions for a variety of system Hamiltonians and initial states, and in
two experimental situations with time reversal and time evolution in one
direction. For this purpose, we evaluate the time dependent overlap of the wave
functions or the fidelity decay during the quantum state
evolutions~\cite{haa18}. In the first situation with time reversal, we exploit
the LE~\cite{Per84, Gou12,Gor06}, which provides a direct measure
of fidelity decay when the high levels are taken and not taken into account. In
the second situation with time evolution in one direction, we evaluate the
overlap of the evolving wave functions considering hard-core bosons with two
energy levels and soft-core bosons with additional higher energy levels. Our
approach quantifies the fidelity decay as a result of population leakage during
state evolution in a QQD experiment. The decay rate is shown to depend on the
ratio of the on-site interaction over coupling strength, the filled particle
number and locations of the initial states, and the applied driving, which is
useful for assessing the two-level approximation of transmons in the QQD
experiments for a number of spin and hard-core boson modelled systems.

Below we first introduce the basic concept of LE and fidelity, and describe the
models of a superconducting $L$-qubit chain ($L$ being the qubit number) taking
into account the external drivings often used in experiments. Then we present
the numerical results of fidelity decay in three cases, those with Floquet
driving,  without driving, and with applied transverse field. The results will
be presented for an $L$ = 10 system in two separate sections for the situations
with and without the time reversal process. All calculations are carried out
considering the initial states and model parameters typically seen in current
realistic experiments. These will be followed by the discussion of the basic
processes in QQD experiments and comparisons with the experimental observations
using superconducting transmon qubits. The final section concludes with a
summary and outlook.

\section{Model and method}

\begin{figure}[t] 
	\includegraphics[width=0.42\textwidth]{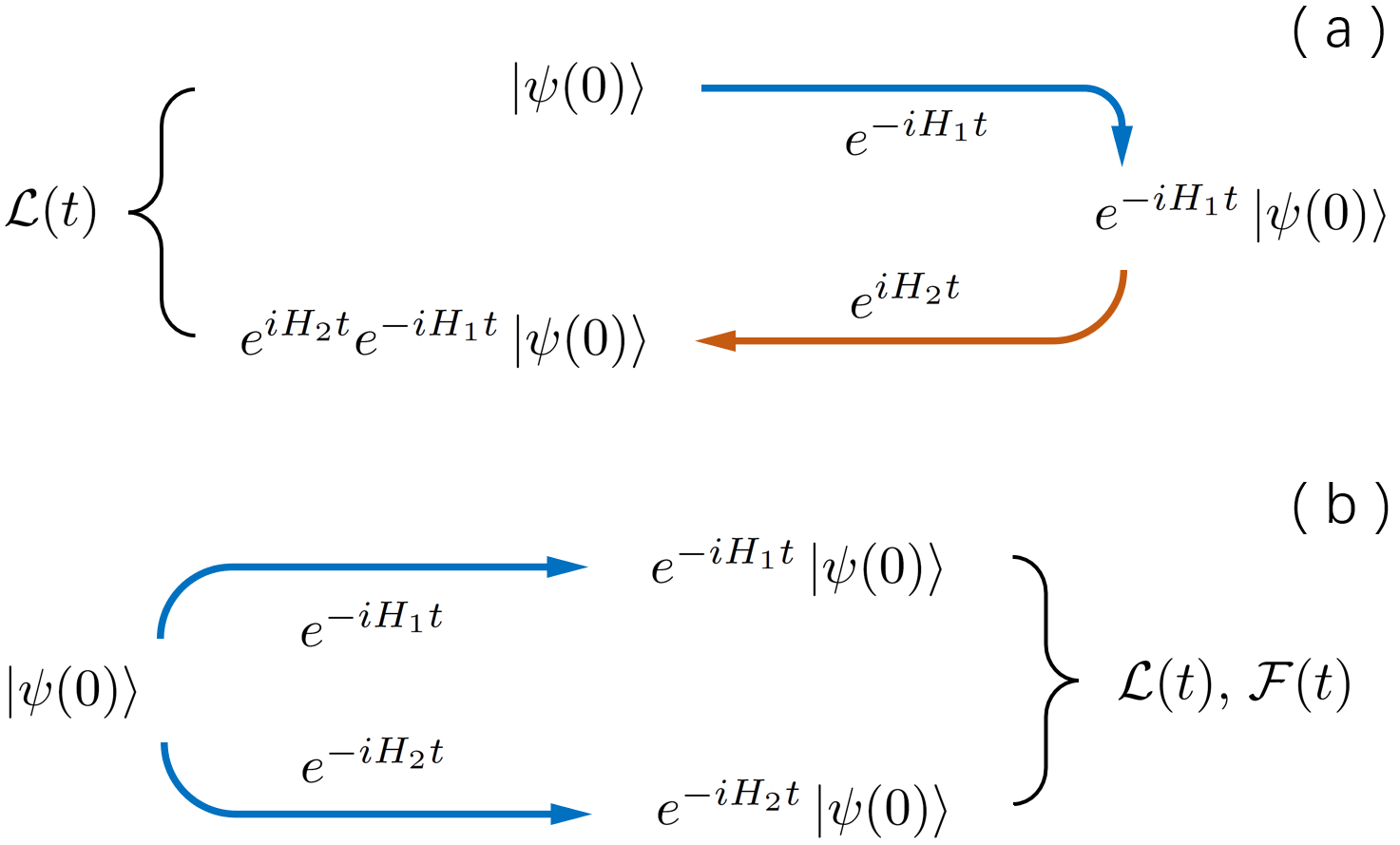} 
	\caption{(a) $\mathcal{L}(t)$ defined as the overlap of the initial state
		$\ket{\psi(0)}$ with the state after the forward evolution under $H_1$ followed
		by the backward evolution under $H_2$ (Loschmidt echo). (b)  $\mathcal{L}(t)$
		obtained from the overlap of states of the forward evolutions from
		$\ket{\psi(0)}$ with $H_1$ and $H_2$, respectively (fidelity).}
	\label{fig1}
\end{figure}

The LE is traditionally a measure of quantum-state revival with imperfect
time-reversal~\cite{Per84, Gou12, var17, has21} and can serve as a benchmark for
the reliability in quantum information processing~\cite{Gor06, jal01, zan12}. It
is defined as the overlap of the initial state $\ket{\psi(0)}$ with the final
state $\ket{\psi(t)}$ obtained after a {\it forward} evolution to time $t$ under
Hamiltonian $H_1$ followed by a {\it backward} evolution from $t$ to 2$t$ under
Hamiltonian $H_2$ [see Fig.~\ref{fig1}(a)]:
\begin{align}
	\mathcal{L}(t) =
	|\bra{\psi(0)}e^{iH_2t}e^{-iH_1t}\ket{\psi(0)}|^2 ~. 
	\label{LE} 
\end{align} 
The existing difference $\Delta H$ = $H_1-H_2$ between $H_1$ and $H_2$ gives
rise to an imperfect recovery of the initial state and therefore
$\mathcal{L}(t)$ is typically a decreasing function of $t$. It describes how
precisely a quantum state can be recovered under a perturbed time reversal
\cite{san20, Zhao22}. It is easy to see that the LE defined in Eq.~(\ref{LE})
can also be interpreted as the overlap of two states obtained after two
\textit{forward} evolutions with different Hamiltonians $H_1$ and $H_2$ [see
Fig.~\ref{fig1}(b)]. In this case, $\mathcal{L}(t)$ is simply a time-dependent
overlap of the two wave functions, usually called fidelity $\mathcal{F}(t)$,
which is used to characterize chaotic and regular motions and the accuracy of
quantum computation in the presence of imperfections \cite{haa18,geo01,gar08}.

In the rotating frame with a common  frequency, the one-dimensional $L$-qubit
system studied in this work is governed, when including external drivings, by
the 1D Bose-Hubbard model~\cite{Roushan2017,Yan2019}:
\begin{equation}
	H(t) = H_0 + H_U + \sum_{j=1}^{L} \omega_{ j}(t)n_{j} + 
	\frac{1}{2}\sum_{j=1}^{L} \Omega_{ j}(a^{\dagger}_{j}+a_{j})~,~
	\label{Hxy} 
\end{equation} 
in which $H_0$ = $\sum_{j=1}^{L-1}J_{j,j+1}(a^{\dagger}_{j} a_{j+1}+\text{h.c.})$
is the hopping term with nearest neighbour coupling strength $J_{j,j+1}$ and
$H_U$ = $\sum_{j=1}^{L}\frac{-U_j}{2} n_{j}(n_{j}-1)$ with the on-site
interaction or anharmonicity $-U_j \le 0$. $a^{\dagger}_{j}$ ($a_{j}$) is the
bosonic creation (annihilation) operator and $n_{j}$ = $a^{\dagger}_{j}a_{j}$ is
the number operator. The Hamiltonian in Eq.~(\ref{Hxy}) includes a term of local
transverse field with strength $\Omega_j$ written in the rotating frame of the
driving frequency resonant with the qubit common frequency. The qubit frequency
can also be tuned by ac magnetic flux in the form $\omega_{ j}(t) =\varepsilon_j
\cos(\nu t)$, with $\nu$ and $\varepsilon_j$ being the frequency and amplitude,
respectively. In this case, without transverse field, $H(t)$ describes a Floquet
system satisfying $H(t)=H(t+T)$ with a period of $T=2\pi/\nu$ ~\cite{eck17}. In
this work, we will consider three cases, namely, quenches with Floquet driving,
without driving, and with transverse field, respectively. We will use constant
coupling and field strengths $J$ and $\Omega$, and varying anharmonicity $U_j$
for individual qubits taken from experiments~\cite{Zhao21,Zhao22}, while the
qubit next nearest neighbour coupling will be neglected (see the Supplemental
Material~\cite{SM}).

Note that the on-site interaction term $H_U$ will be absent if the system is
populated only with one particle and the noncomputational high energy levels are
not involved. The central goal of this work is to see the difference between the
evolving quantum states with and without considering $H_U$ when high levels are
only sparsely populated. For the experiment with time reversal, we can reverse
the time evolution of the system in~Eq.(\ref{Hxy}) by changing the signs of the
coupling $J$, either directly via tunable coupler~\cite{yan18} or by Floquet
driving~\cite{Zhao21,Zhao22}, and of the transverse field strength $\Omega$ by
tuning the phase of driving field~\cite{guo18,Xu2020}, so that all terms in $H$
except $H_U$ change their signs. Therefore, $\mathcal{L}(t)$ describes the
fidelity decay in a time reversed experiment that would be ignored if the
two-level model is used, since in the model $H_U$ is absent giving rise to a
perfect time reversal with fidelity of unity. 

Without loss of generality, we consider the simple case in the absence of external 
drivings (Floquet and transverse field). The Loschmidt echo can be written as 
$\mathcal{L}(t)$ = $|\bra{\psi(0)}e^{i(H_0-H_{U})t}e^{-i(H_0+H_{U})t}\ket{\psi(0)}|^2$.
The difference between the Hamiltonians of the forward (from $0$ to $t$) and
backward (from $t$ to $2t$) evolution is $\Delta H$ = 2$H_{U}$. As explained
above, by calculating Loschmidt echo $\mathcal{L}(t)$ in Fig.~\ref{fig1}(a), we
identically obtain the fidelity between two states evolving with Hamiltonians
$H_{1,2}$ = $H_0 \pm H_U$ as illustrated in Fig.~\ref{fig1}(b). Since the
timescale under consideration is well below the experimentally achievable
coherence times (see SM~\cite{SM}), dissipations are ignored in this work and
evolutions of pure states are used in the numerical simulation. Namely, the
Loschmidt echo $\mathcal{L}$ or the fidelity $\mathcal{F}$ can be calculated
with pure states by $\mathcal{F}(t)  = |\langle
\psi(0)|\psi(t)\rangle|^2$, in which $\ket{\psi(t)}$ is the final
state at $2t$. Or equivalently, $\mathcal{F}(t)  = |\langle
\psi_1(t)|\psi_2(t)\rangle|^2$, where $\ket{\psi_{1,2}(t)}$ =
$e^{-iH_{1,2}t}\ket{\psi(0)}$~\cite{nie10}.

For the experiment with time evolution in one direction, we identically
calculate $\mathcal{L}(t)$ for a forward evolution to time $t$ from a given
initial state under $H_0$ in the Fock state site basis $\ket{n_1,n_2,...,n_{L}}$
with $n_i$ = 0, 1, and a backward evolution under $H_1$ = $H_0+H_U$ in the basis
$\ket{n_1,n_2,...,n_{L}}$ with $n_i$ = 0, 1, ..., $K-1$ ($K$ being the number of
qubit levels considered). The basis for the states in the forward
evolution is expanded to that in the backward evolution for the calculation.
This is again equivalent to the time dependent overlap or fidelity decay between
the two wave functions evolving from the same initial state to time $t$ under
the two different Hamiltonians $H_0$ and $H_1$, which offers a proper estimate
of the error in the two-level approximation of transmons for experiments with
time evolution in one direction.

\section{Numerical results with time reversal}\label{TR}

In this section, we present the results for a 10-qubit chain ($L$ = 10) with
time reversal process in three different cases using typical experimental
parameters of superconducting transmon qubits (see the Supplemental
Material~\cite{SM}). The calculations are performed using the QuTiP
toolbox~\cite{Joh12,Joh13} considering three energy levels ($K$ = 3) for each
qubit. The high level therefore refers to the energy level of the qubit
second-excited state. We will use the terms interchangeably. For more
straightforward and convenient comparisons with the reported experiments, the
time evolutions will be shown in nanoseconds and the normalized time $tJ/2\pi$
is used in the place required with which $tJ/2\pi$ = 1 corresponds to the time 
$t$ = 1 $\mu$s when $J/2\pi$ = 1 MHz.

\subsection{Quench with Floquet driving}


\begin{figure}[t] \includegraphics[width=0.36\textwidth]{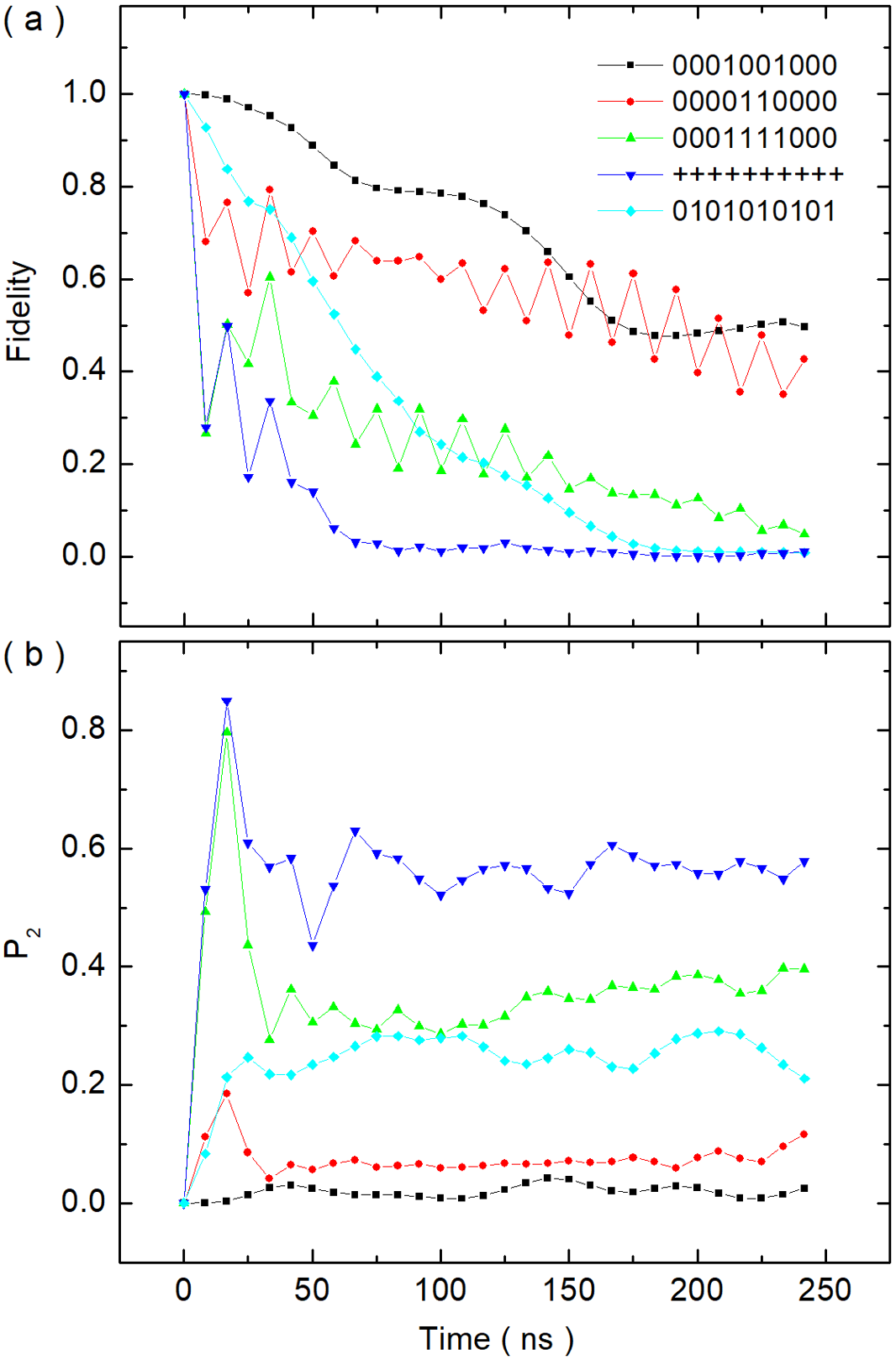}
	\caption{Time dependence of (a) fidelity and (b) total population of the second
		excited state during forward evolution. The initial states are $\psi_1$ =
		$\ket{0001001000}$, $\psi_2$ = $\ket{0000110000}$, $\psi_3$ =
		$\ket{0001111000}$, $\psi_4$ = $\ket{++++++++++}$, and $\psi_5$ =
		$\ket{0101010101}$. Time reversal is performed by Floquet driving. See text for
		the parameters used in the numerical calculations.}
	\label{fig2}
\end{figure}

Periodically modulating many-body systems using sinusoidal or non-sinusoidal
drive is a standard tool for the studies of gauge fields, topological band
structures, phase transition~\cite{eck17}, and nonequilibrium state of matter
such as the discrete time-crystalline phase~\cite{Mi22}. In Eq.~(\ref{Hxy}),
with $\nu \gg J$ and  in the absence of $H_U$ and transverse field, we 
obtain a time-independent Hamiltonian $H_{\text{eff}}$ =
$\sum_{j=1}^{9}J_{\text{eff}}(a^{\dagger}_{j} a_{j+1}+\text{h.c.})$ with
$J_{\text{eff}} \approx J\mathcal{J}_0(\varepsilon/\nu)$, where
$\mathcal{J}_0\big(x\big)$ is the Bessel function of order zero~\cite{eck17}.
This is realized by driving the odd-number qubits with the same amplitude
$|\varepsilon_j|=\varepsilon$ and staggered phase~\cite{Zhao21, Zhao22}.
$J_{\text{eff}}$ can thus be tuned from positive to negative by changing
$\varepsilon$ and $\nu$ resulting in a time-reversible system $H_{\text{eff}}$.

In Fig.~\ref{fig2}, we show the calculated time dependence of fidelity and total
population of the qubit second-excited state $P_2$ = $\sum_{j=1}^{10}P_2^j$
($P_2^j$ for the $j$th qubit) during forward evolution for five different
initial states of $\psi_1$ = $\ket{0001001000}$, $\psi_2$ = $\ket{0000110000}$,
$\psi_3$ = $\ket{0001111000}$, $\psi_4$ = $\ket{++++++++++}$, and the N\'{e}el
state $\psi_5$ = $\ket{0101010101}$ ($\psi_i$ stands for $\ket{\psi_i(0)}$).
Here $\ket{0}$ and $\ket{1}$ are the qubit ground and first-excited states, and
$\ket{+}$ is the eigenstate of Pauli matrix $\sigma^x$ with eigenvalue of $+1$.
In the calculations, we set $J/2\pi$ = 10.8 MHz and $U_j/2\pi$ around 212 and
264 MHz for odd and even number qubits, respectively~\cite{SM}. Also, we fix
$\nu/2\pi$ = 120~MHz, and the driving amplitude is $\varepsilon_f/2\pi=
213.6$~MHz and $\varepsilon_b/2\pi= 400$~MHz for the forward and backward
evolutions, which lead to $\mathcal{J}_0(\varepsilon_f/\nu)
=-\mathcal{J}_0(\varepsilon_b/\nu)$ corresponding to $J_{\text{eff}}/2\pi$
$\approx\pm 3.8$~MHz. All the data are collected in a stroboscopic fashion in
the step of the driving period $2\pi/\nu$ $\approx$ 8.33 ns~\cite{eck17}.

We choose five different initial states to see their different influence on the
fidelity decay. $\psi_1$ and $\psi_2$ have the total particle number $N$ = 2
(note $N$ is conserved), but they have different particle distributions.
$\psi_3$, $\psi_5$, and $\psi_4$ have $N$ = 4, $N$ = 5, and an expectation value
of 5, respectively. In Fig.~\ref{fig2}, we find a general trend that fidelity
decays faster with the increase of $N$ that results in the increase of high
level excitations. An initially fast rising of the excitations leads to fast
dropping of the fidelity. In addition, the spatial separation of particles will
cause the process less abrupt, as can be seen from the comparison of the data
between $\psi_1$, $\psi_5$ and $\psi_2$, $\psi_3$, $\psi_4$. Detailed results of
the population variations with time for individual qubits, during both the
forward and backward evolutions can be found in the Supplemental
Material~\cite{SM}. There it is seen that in the cases of $\psi_2$ and $\psi_3$
the populations in the qubit high level tend to remain in the initially excited
qubits which locate next to each other. This is not the case when they are
separated, as in the case of $\psi_1$.

\subsection{Quench without driving}

\begin{figure}[t] 
	\includegraphics[width=0.38\textwidth]{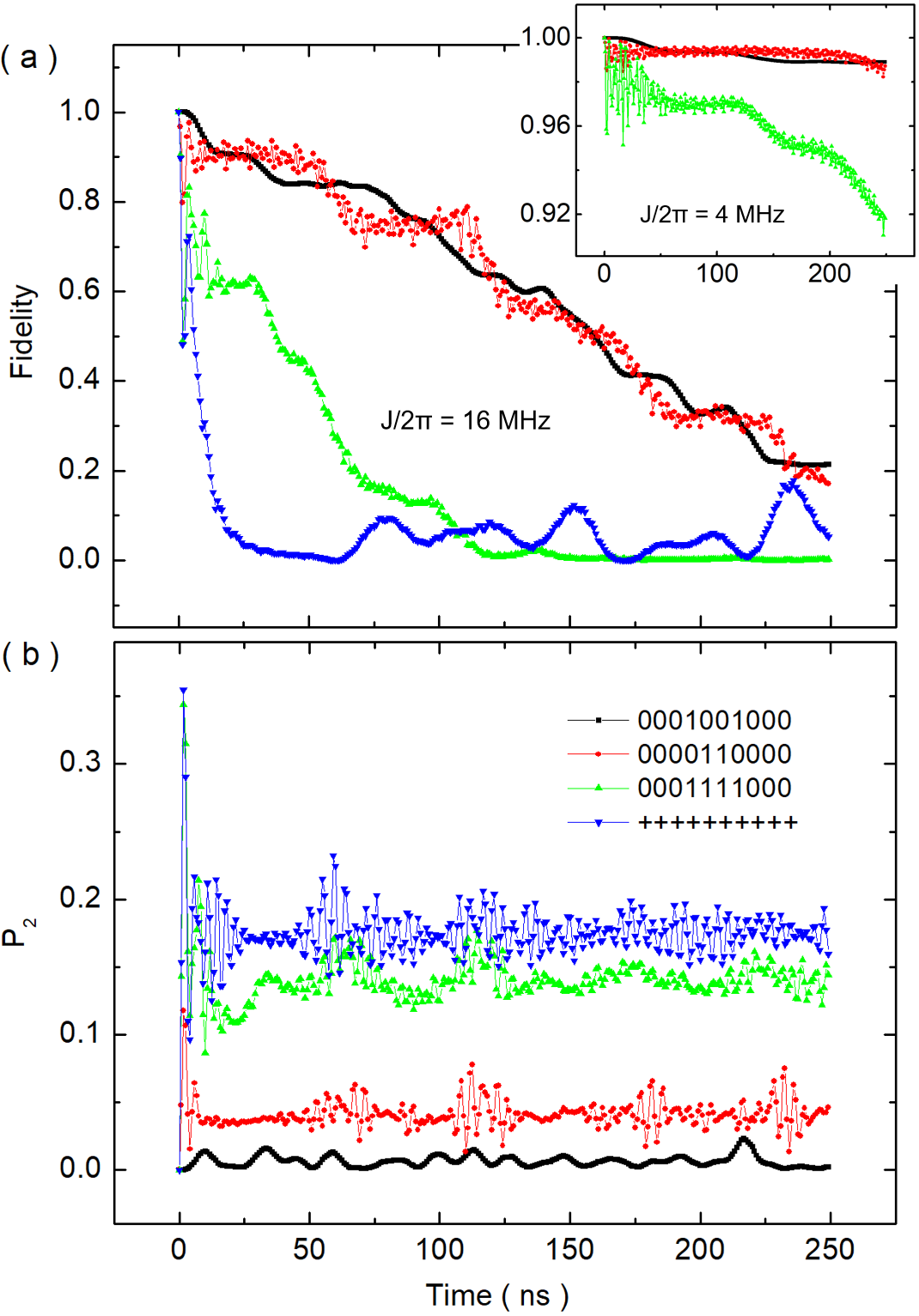} 
	\caption{Time dependence of (a) fidelity and (b) total population of the second
	excited state during forward evolution without external driving. The initial
	states are $\psi_1$ = $\ket{0001001000}$, $\psi_2$ = $\ket{0000110000}$,
	$\psi_3$ = $\ket{0001111000}$, and $\psi_4$ = $\ket{++++++++++}$. Time reversal
	is performed by changing the sign of the coupling strength $J$, which is 2$\pi
	\times$16 and 2$\pi \times$4 MHz in the main panels and inset, respectively.}
	\label{fig3} 
\end{figure}

Figure~\ref{fig3} shows the results for the same initial states in the cases
without Floquet driving and transverse field. The time reversal is implemented
by directly changing the sign of qubit coupling $J$ in the numerical
calculations. In the main panels with $J/2\pi$ = 16 MHz, one can again see the
more significant fidelity decay with increasing $N$, but the fidelities in the
cases of $\psi_1$ and $\psi_2$ become almost the same although their high-level
excitations are clearly different. The difference as compared to the results in
Fig.~\ref{fig2} may result from some detailed processes which are presently
unclear. In the Supplemental Material~\cite{SM}, it can also be seen that the
populations in the qubit high level tend to remain in the initially excited
qubits when located next to each other.

\begin{figure}[t] \includegraphics[width=0.38\textwidth]{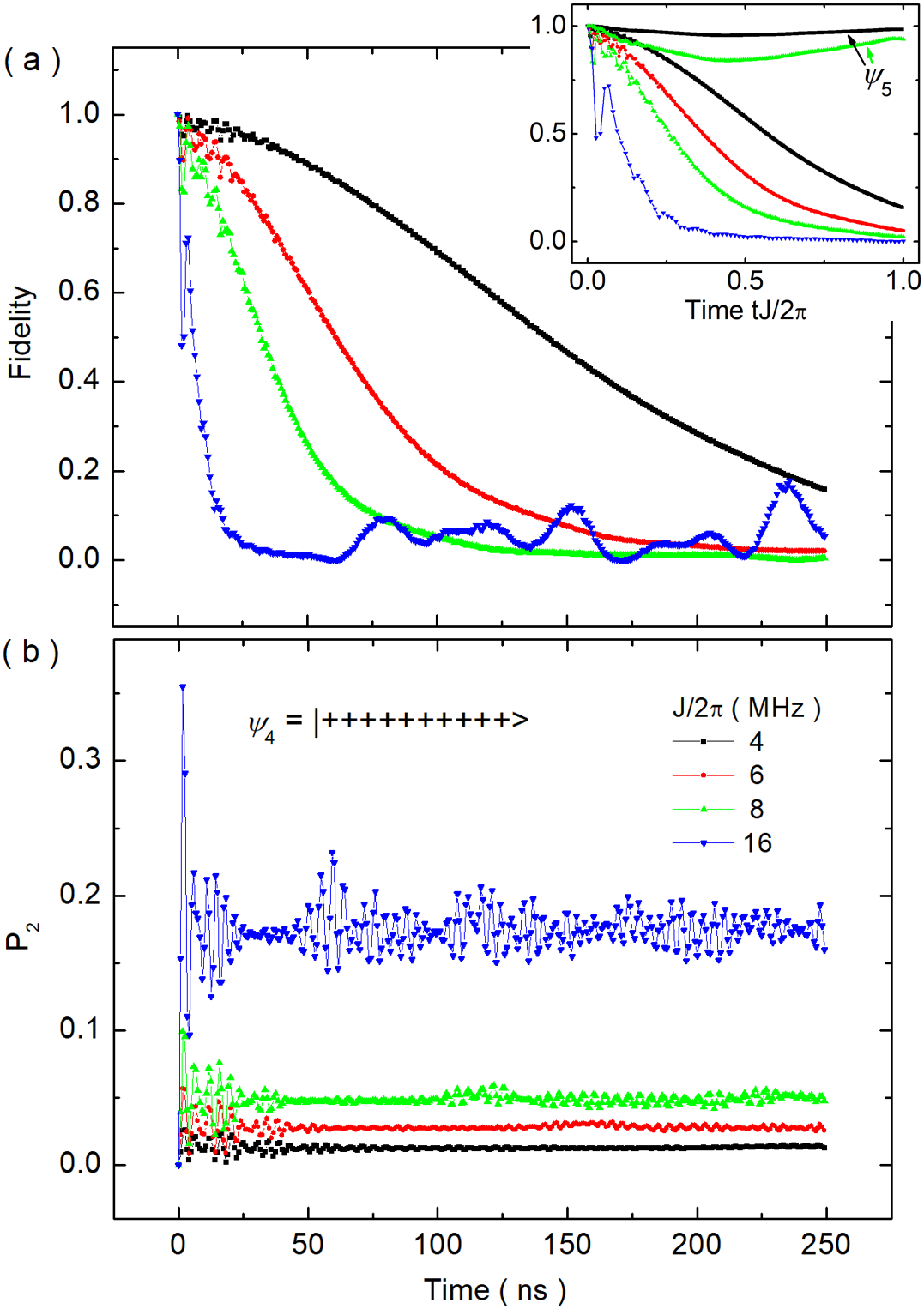} 
	\caption{Time dependence of (a) fidelity and (b) total population of the second
	excited state during forward evolution without external driving. The initial
	state is $\psi_4$ = $\ket{++++++++++}$ and different $J/2\pi$ = 4, 6, 8, and 16
	MHz are used. Time reversal is performed by changing the sign of $J$. The inset
	shows the fidelity against time $tJ/2\pi$ with two additional results for
	N\'{e}el state $\psi_5$ = $\ket{0101010101}$ with $J/2\pi$ = 4 and 8 MHz.}
    \label{fig4}
\end{figure}

An important observation from the numerical simulations is that the fidelity
decay strongly depends on the qubit coupling strength $J$ as well as on the
configuration of the initial state. This is expected since reducing $J$ leads to
an increasing ratio $U/J$ while the hard-core boson model is arrived when
$U/J$ $\rightarrow$ $\infty$~\cite{Caz11}. The inset of Fig.~\ref{fig3} shows
the results for $\psi_1$, $\psi_2$, and $\psi_3$ with a reduced $J/2\pi$ of 4
MHz, as compared to the data with $J/2\pi$ = 16 MHz in the main panels. The
corresponding results for $\psi_4$ are presented separately in
Fig.~\ref{fig4}(a). In the inset of Fig.~\ref{fig3}, the fidelity only shows a
slight decrease for $\psi_1$ and $\psi_2$, a moderate decrease for $\psi_3$, but
a significant decay for $\psi_4$ as shown in Fig.~\ref{fig4}(a). In the cases of
$\psi_2$ and $\psi_3$, the populations at high level are on the order of 0.1$\%$
and 0.2$\%$ for the initially excited qubits, respectively, and the values for
the other qubits and those for all qubits in the case of $\psi_1$ are
negligible. Looking at the populations in the first-excited state of the
initially excited qubits in these cases, we find that they almost return to
unity ($\sim$0.994, 0.992, and 0.974 for $\psi_1$, $\psi_2$, and $\psi_3$,
respectively) at the end of the time reversal process (see the Supplemental
Material~\cite{SM}).

The fidelity decay and total population versus time for several different $J$
are presented in Fig.~\ref{fig4} for the case of $\psi_4$. The results
demonstrate a fast decay rate increase as $J$ increases. To subtract the factor 
of the increase of $J$, which measures the time of hopping process, the decay
rate increase is still substantial, as can be seen in the inset of
Fig.~\ref{fig4} plotted against $tJ/2\pi$. These results indicate strong influence of
the qubit coupling strength on the fidelity decay in the QQD experiments. As can
be seen in Figs.~\ref{fig2} and \ref{fig3}, the decay rate also depends on the
initial state. In the inset of Fig.~\ref{fig4}, we show two additional results
for the N\'{e}el initial state $\psi_5$ = $\ket{0101010101}$ with $J/2\pi$ = 4 (solid line)
and 8 (dashed line) MHz. It can be seen that for the N\'{e}el state with qubits
initially excited on every other site, the decay rate decreases substantially in
the regime of small $J$ and the state leakage to the qubit high level becomes
negligible.

Comparing the data in Fig.~\ref{fig2} with $J/2\pi$ = 10.8 MHz and
$J_{\text{eff}}/2\pi$ $\approx 3.8$~MHz to those in Figs.~\ref{fig3} and
\ref{fig4}, we find that in the case of Floquet driving, the degrees of the
fidelity decay and high-level population are more related to the qubit actual
coupling strength than to the effective one, since for the case without driving
the coupling strength $J/2\pi$ = 4 MHz leads to a fidelity decay much 
less severe.

\subsection{Quench with transverse field}

\begin{figure}[t] \includegraphics[width=0.38\textwidth]{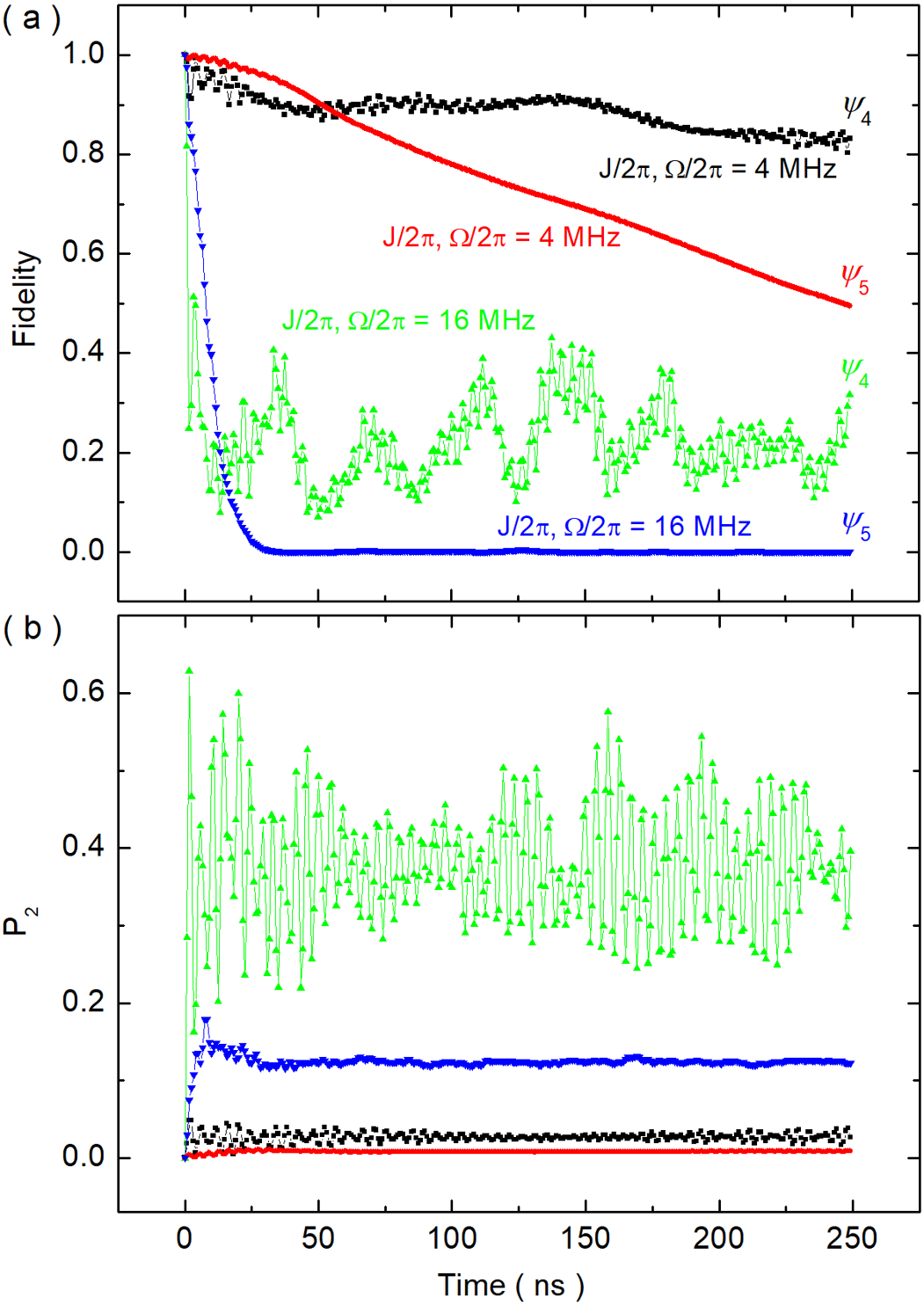} 
	\caption{Time dependence of (a) fidelity and (b) total population of the second
	excited state during forward evolution with applied transverse field and two
	initial states $\psi_4$ = $\ket{++++++++++}$ and $\psi_5$ = $\ket{0101010101}$
	for different coupling and field strengths $J$ and $\Omega$. Time reversal is
	performed by changing the signs of $J$ and $\Omega$.}
	\label{fig5} 
\end{figure}

The application of a transverse field leads to the last term in Eq.~(\ref{Hxy})
and changes the system from integrable to nonintegrable. The particle
conservation is broken in this case. These give rise to many interesting
phenomena which have been studied recently with superconducting
circuits~\cite{Xu2020,Chen2021}. In Fig.~\ref{fig5}, we show the time
dependence of fidelity and total population of the second-excited state for two
initial states $\psi_4$ and $\psi_5$ with two different sets of coupling
strength $J$ and transverse field $\Omega$. Again we see that the fidelity decay
rate decreases quickly with decreasing $J$. The results also show opposite
trends of decay rates for $\psi_4$ and $\psi_5$ as compared to the data in
Fig.~\ref{fig4}. Namely, although the populations in the second-excited state
are larger for the $\psi_4$ state than that for the $\psi_5$ state with the same
$J$ and $\Omega$, the fidelity decays more significantly for the N\'{e}el state
$\psi_5$ than for the $\psi_4$ state. These are apparently contrary to 
the results in Fig.~\ref{fig4}.

\begin{figure}[t] \includegraphics[width=0.38\textwidth]{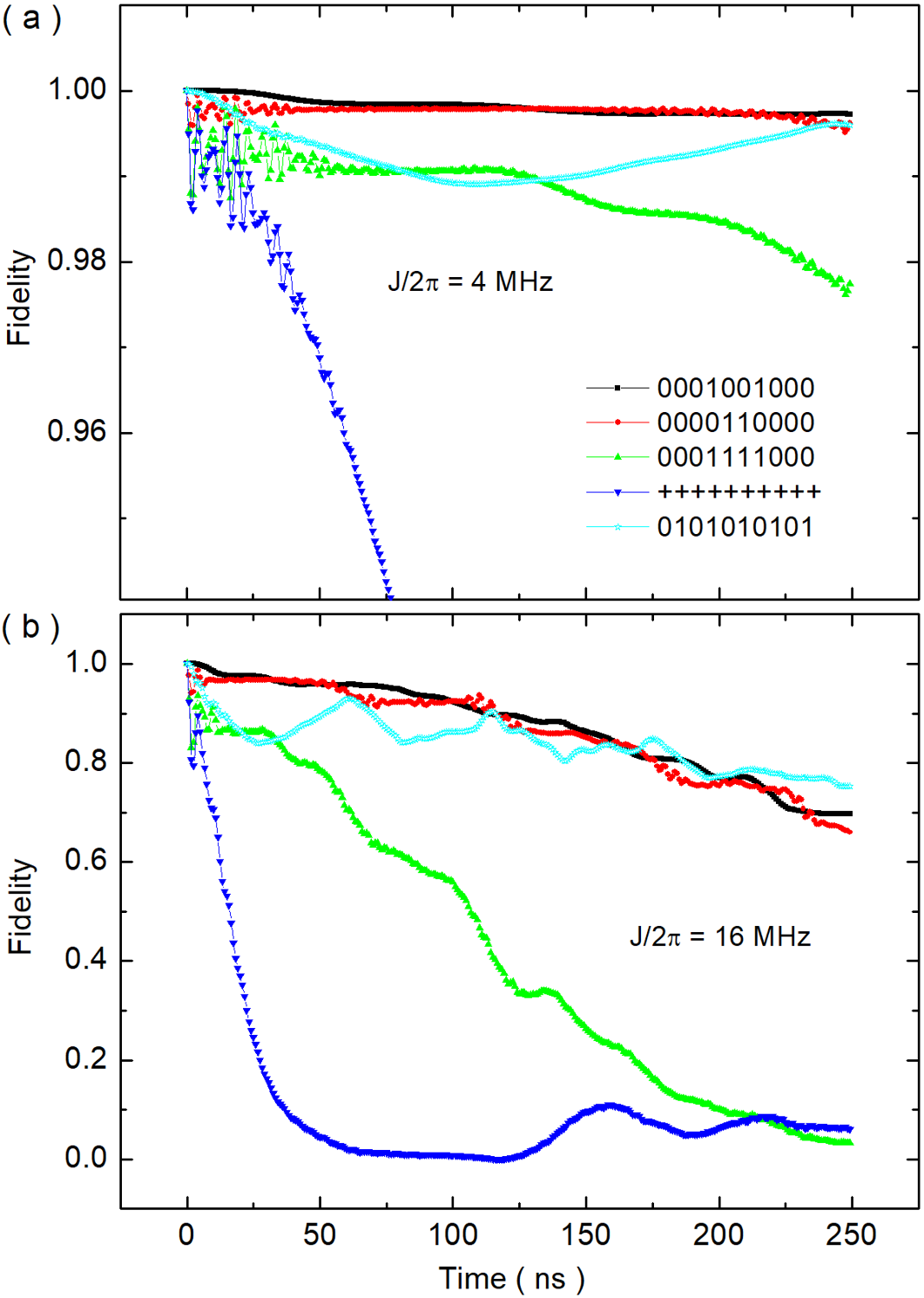} 
	\caption{Time dependence of fidelity with (a) $J/2\pi$ = 4 MHz and (b) $J/2\pi$
	= 16 MHz for the initial states $\psi_1$ = $\ket{0001001000}$, $\psi_2$ =
	$\ket{0000110000}$, $\psi_3$ = $\ket{0001111000}$, $\psi_4$ =
	$\ket{++++++++++}$, and $\psi_5$ = $\ket{0101010101}$. The evolutions are in a
	single forward direction without external driving.}
	\label{fig6} 
\end{figure}

The fact that the decay rate becomes higher for $\psi_5$ than that for $\psi_4$
may relate in a complicated way to the interplay of the high-level excitation
and the well-known process of thermalization in nonintegrable chaotic systems
with applied transverse field~\cite{ban11,Chen2021}. Specifically, for the
results in Fig.~\ref{fig5} the system can be identified as chaotic, and strong
and weak thermalizations occur for the initial states of $\psi_5$ and $\psi_4$,
respectively. In the Supplemental Material~\cite{SM}, we show their distinct
behaviours of the entanglement entropy in the forward and backward evolution
directions.  Also, we find that for the N\'{e}el state $\psi_5$ with $J/2\pi$ =
$\Omega/2\pi$ = 16 MHz, the populations of the first-excited state $P_1$ for
each qubit will oscillate and approach around 0.5 in a timescale of $\sim$100
ns. An interesting observation is that the populations of each qubit stay around
0.5 afterwards and do not show any sign of changing back even when the time
reversal process is performed. A population of 0.5 corresponds to a zero
expectation value of $\sigma^z$.  In this strongly thermalized regime, the
expectation values of other Pauli matrices will only have small-magnitude
oscillations around zero. This means that the Bloch vector of each qubit
progressively shrinks in the thermal state and does not change back
afterwards~\cite{rem1}. Such a feature of local operators is consistent with
that of the entanglement entropy in the Supplemental Material~\cite{SM}.
Considering the identical view in Fig.~\ref{fig1}(b) with two wave functions
evolving in the same forward direction, the results imply that thermalization
generally aggravates the fidelity decay in the QQD experiment with high-level
participation.

Without high-level populations, the system is described by a two-level model and
can have a perfect time reversal with unity fidelity also in the thermalized
state. For the N\'{e}el state with $J/2\pi$ = $\Omega/2\pi$ = 4 MHz and small
high-level populations, partly reversed processes are found from population
$P_1$ as well as entanglement entropy.  The situation is quite different in the
weakly thermalized regime for the $\psi_4$ initial state, as indicated by the 
entanglement entropy in the Supplemental Material~\cite{SM}. The expectation
value of $\sigma^x$ will oscillate around finite values while those of the other 
Pauli matrices around zero with moderate amplitudes. Compared with the data 
in Fig.~\ref{fig4}, the results in Fig.~\ref{fig5} shows larger high-level populations 
but a smaller fidelity decay for the initial state $\psi_4$, which remains to be
explained.

\section{Numerical results with time evolution in one direction}

\begin{figure}[t] \includegraphics[width=0.38\textwidth]{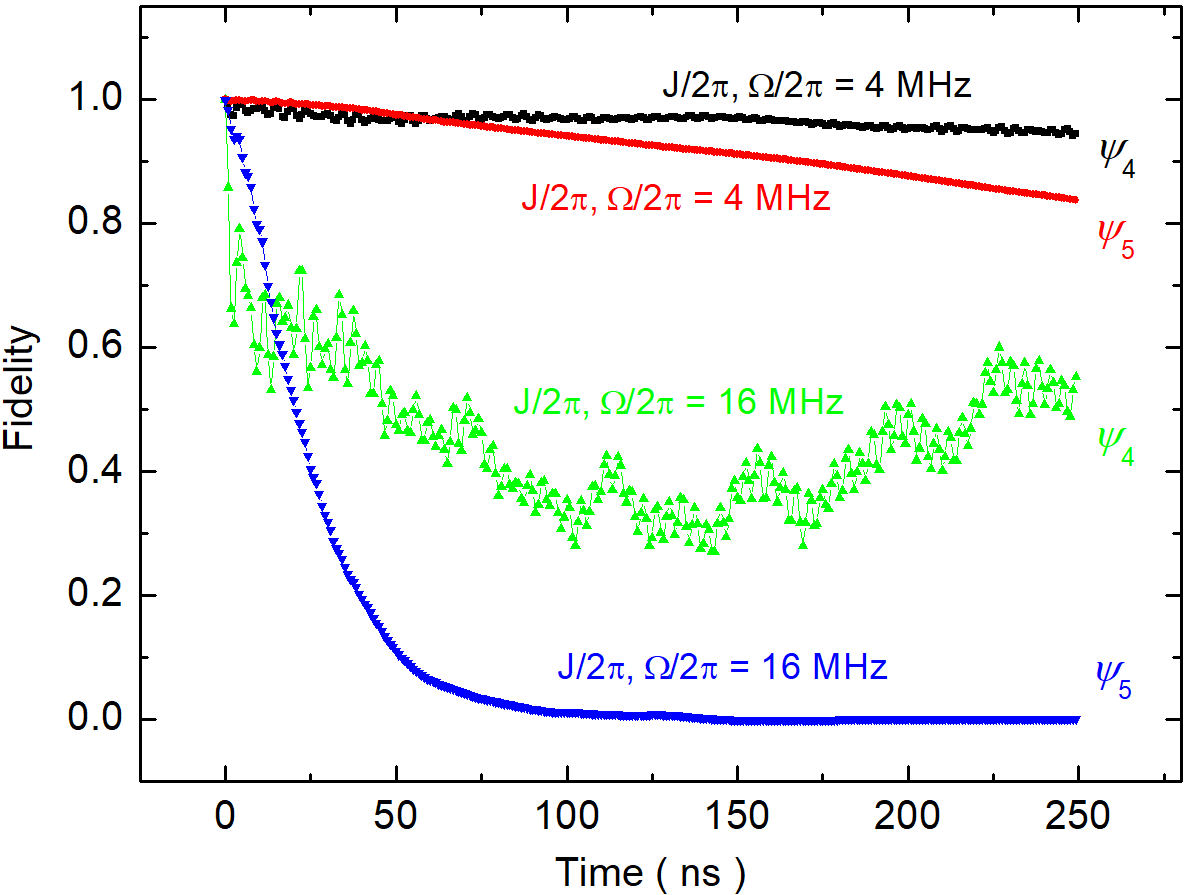} 
	\caption{Time dependence of fidelity with applied transverse field and two
	initial states $\psi_4$ = $\ket{++++++++++}$ and $\psi_5$ = $\ket{0101010101}$
	for different coupling and field strengths $J$ and $\Omega$. The evolutions are
	in a single forward direction.}
	\label{fig7} 
\end{figure}

\begin{figure*}[t] \includegraphics[width=0.99\textwidth]{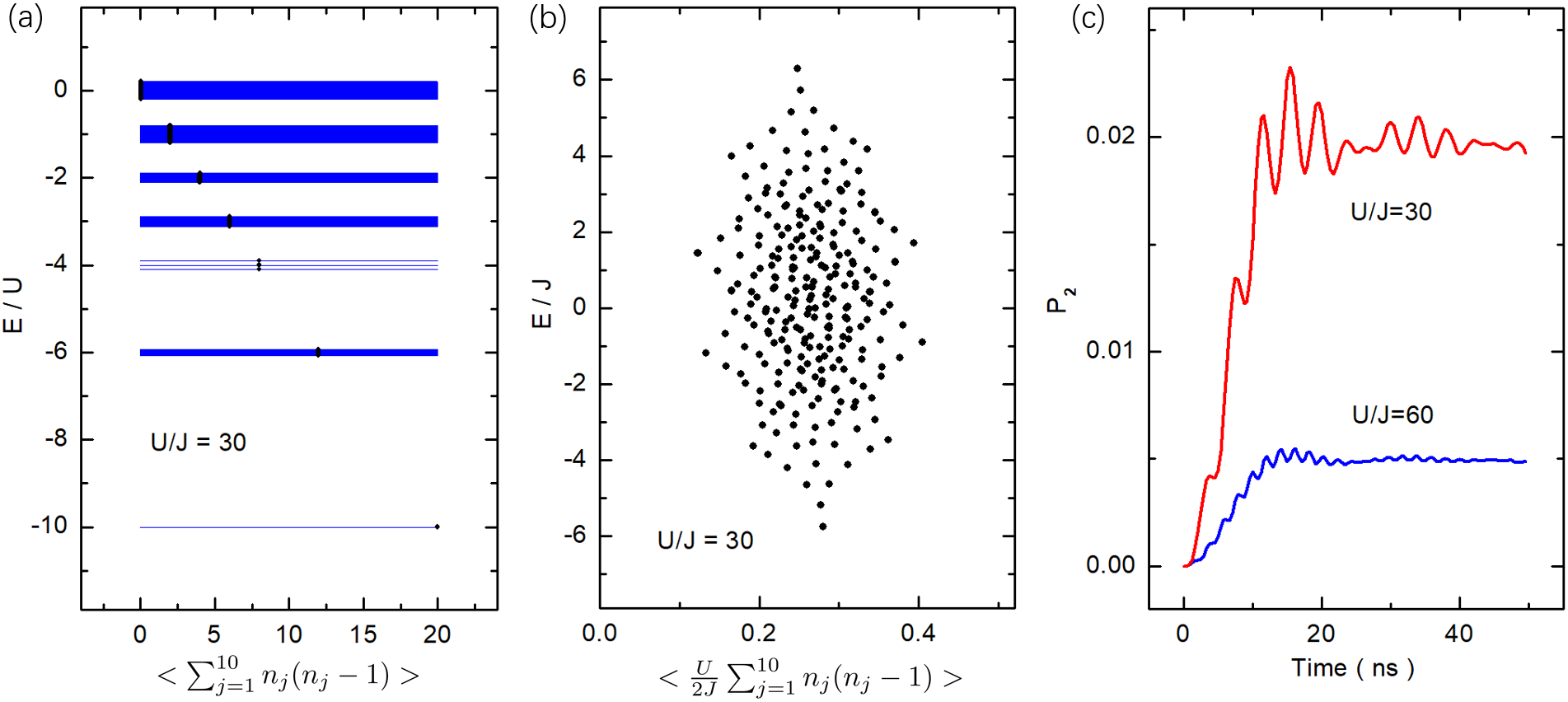} 
	\caption{Simple particle conserved system $H$ = $H_0 + H_U$ of the 10-qubit
		chain. (a) Energy levels forming bands with $U/J$ = 30 and $N$ = 5. The black
		dots show the expectation values $A$ of the anharmonicity operator
		$\sum_{j=1}^{10}n_{j}(n_{j}-1)$. (b) Enlarged plot of the expectation values of
		the anharmonicity operator in the top band showing the eigenvalues in unit of
		$J$ against $(U/2J)A$. (c) Total population of the second-excited state in the
		transitional stage right after quenching of the system for the N\'{e}el initial
		state $\psi_5$ with $J/2\pi$ = 8 MHz, and $U/2\pi$ = 240 (upper curve) and 480
		(lower curve) MHz.}
	\label{fig8} 
\end{figure*}

Many QQD experiments using superconducting circuits are performed with quantum
state evolution in a single forward direction~\cite{Xu2018, Yan2019, Xu2020,
	Gong2021, Chen2021, Yanay20, Kara2022}. In Figs.~\ref{fig6} and \ref{fig7}, we
show the numerical results without the time reversal process, which correspond
to the data in Figs.~\ref{fig3}, \ref{fig4}, and \ref{fig5} with time reversal.
It can be seen that the general trends are the same for the two situations. In
particular, for larger coupling strength $J/2\pi$ = 16 MHz, the fidelity decay
rates are almost identical if twice the total time duration in the time reversed
situation is considered. However, with $J/2\pi$ = 4 MHz, the fidelity appears to
decay faster in the time reversed situation after considering the twice time
duration. This is generally understandable since in this case the two quantum
states evolve from the same initial state under $H_1$ and $H_2$ (or both plus
the transverse field term) with different terms $\pm H_U$ of opposite signs.
Therefore the fidelity decay rate is larger compared to the situation with time
evolution in one direction in which the states evolve under $H_0$ and $H_1$ (or
both plus the transverse field term) with the term $H_U$ appearing only in
$H_1$.

\section{Discussions}

In a QQD experiment, individual qubits in the multiqubit system are first biased
at their respective idle frequencies, where they are essentially decoupled, to
prepare a ground, excited, superposition, or entangled state. With these
constituting the initial state, the qubits are biased at the quench time $t$ = 0 to
the same resonant working frequency at which a particular Hamiltonian $H(t)$ in
Eq.~(\ref{Hxy}) is used for the system evolution. Finally, all qubits are set 
back to their idle frequencies and the state of each qubit is measured tomographically.
Physically, before $t$ = 0, the decoupled qubit system can be described via
the Fock state site basis $\ket{n_1,n_2,...,n_{L}}$ with a Hilbert space dimension of
$K^{L}$ when the number of levels $K$ is considered for each qubit. However,
only the bottom two levels are used for the initial state construction and all the
high levels are unoccupied. At $t$ = 0, the initial state can be expanded in
terms of the eigenbasis of $H$ with the same Hilbert space dimension. The state
will then evolve according to Schr\"{o}dinger’s equation, producing high-level 
populations which can be measured when the qubits are brought back to the
idle points and the evolved state is expressed in Fock state site basis again.

\subsection{Process at the initial stage after quench}

For the present case of positive $J$ and negative $-U$ in the Hamiltonian $H(t)$
in Eq.~(\ref{Hxy}), the system has mostly negative eigenvalues with zero energy
near the top of the energy spectrum, and the total energy of the initial states
considered in this work is (near) zero. As a concrete example, we consider the
simple particle conserved system $H$ = $H_0 + H_U$ of the 10-qubit chain with
the N\'{e}el initial state and constant $J/2\pi$ = 8 MHz, $U/2\pi$ = 240 MHz for
all qubits corresponding to $U/J$ = 30. For this system, the eigenvalue problem
can be accurately solved in the $N$-particle block of Hamiltonian matrix with
$K$ = $N+1$ levels for each qubit. The block, or the Hilbert space dimension is
$(N+L-1)!/(L-1)!/N!$ for the $L$-qubit chain, which equals 2002 for $L$ = 10 and
the N\'{e}el initial state with $N$ = 5. In Fig.~\ref{fig8}(a), we show the
energy spectrum calculated using the QuSpin package~\cite{Wein17,Wein19}. The
black dots in the figure are the expectation values of the anharmonicity
operator $A$ = $<\sum_{j=1}^{10}n_{j}(n_{j}-1)>$. For the present large value of
$U/J$ = 30, the spectrum clearly shows the band structure with band separation
in unit of $U$ and grouped by different values of $A$. In Fig.~\ref{fig8}(b), we
replot $A$ in the top band in which the lift of the eigenvalue degeneracy on the
order of the coupling strength $J$ can be seen~\cite{Ore19,Man22}.

The upper curve in Fig.~\ref{fig8}(c) shows the temporal variation of the total
population of the second-excited state after the system quench at $t$ = 0, with
the same parameters of $J$ and $U$ described above. The lower curve is the
result with increased $U/2\pi$ = 480 MHz and $U/J$ = 60. We see that doubling
$U/2\pi$ roughly reduces the second-excited state population to a quarter of the
value after the transitional period. In addition, the two curves show well
defined oscillations with a frequency equal to $U/2\pi$, which are more
prominent at the beginning after quench. This can be explained by the results in
Fig.~\ref{fig8}(a). In the $N$ = 5 system, the lowest energies in the figure
correspond to the configurations of all five particles residing in the same
qubit thus leading to $A$ $\sim$ 20. On the other hand, in the top band each
qubit has no more than one particle, while in the second band from the top, only
one qubit is allowed to have increased two particles, leading to the nonzero
population in the second excited state. The total energy of the system is zero
and does not change with time. Therefore, the system has the largest probability
in the eigenstates in the top band and the oscillations in the figure reflect
the separation of the top bands. The corresponding oscillations in fidelity with
the same frequency of $U/2\pi$ as in the second-excited state population, which
can be seen in Figs.~\ref{fig3} to \ref{fig7}, provides a clear  indication that
the fidelity is influenced by the presence of the qubit high level.

We have used three energy levels for each qubit ($K$ = 3) in our numerical
simulations in the previous sections. The results are accurate in the cases of
the initial states $\psi_1$ and $\psi_2$ with $N$ = 2. For the other initial
states, the calculation should be approximate. For the results in
Fig.~\ref{fig8}(a), there will be only three top bands left if we use three
levels for each qubit in stead of six levels in the calculation. We find that
the deviations of both the eigenvalues and expectation values of the
anharmonicity operator such as the data in Fig.~\ref{fig8}(b) are small by using
three and six levels in the calculation, and those for the top band are
smallest. Since for the N\'{e}el initial state the system has zero total energy,
it has the probabilities in the eigenstates orders of magnitude smaller when
going from the top band to the lower bands, so that the level structure of the
top band basically determines the system property during evolution. As a result,
numerical simulations using three levels for each qubit are expected to be a
good approximation.

\subsection{Comparison with experiment}

We now compare our numerical results with some of the QQD experiments using
superconducting transmon qubits~\cite{Xu2018, Yan2019, Xu2020, Gong2021,
	Chen2021, Braum2021, Kara2022, Zhao21, Zhao22}.

Strongly correlated quantum walks have been studied with adjacently and
separately placed two photons in a 12-qubit superconducting
processor~\cite{Yan2019}. The fermionization of strongly interacting photons and
the antibunching of photons with attractive interactions have been observed,
which is described by the hard-core boson model in good approximation.
The superconducting processor has the parameters of $J/2\pi$ $\sim$ 12 MHz, and
two sets of $U/2\pi$ $\sim$ 200 and 240 MHz for the even and odd site qubits,
respectively. The situation is similar to those discussed in Figs.~\ref{fig3}
and \ref{fig6} for the initial states of $\psi_1$ and $\psi_2$ with similar
$U/2\pi$ and an intermediate $J/2\pi$ between 16 and 4 MHz. The populations of
the second-excited state for individual qubits are discussed in
Ref.~\cite{Yan2019}, which can be seen in more detail with present parameters in
the Supplemental Material~\cite{SM} together with the total population in
Fig.~\ref{fig3}(b).

The many-body localization phenomenon arising from the interplay between
interaction and disorder has been observed in a superconducting circuit with
all-to-all connectivity~\cite{Xu2018}. The ordered N\'{e}el initial state
$\ket{0101010101}$ is prepared, and the system is found to have a uniform
population of 0.5 for all qubits at the end of the state evolution in the
absence of disorder. However, in strong disorder the system maintains the
population distribution of the initial state. These experimental results are
found to be well described by the two-level spin model. The superconducting
processor has a very small coupling strength $J/2\pi$ below 2 MHz and an on-site
interaction $U/2\pi$ around 240 MHz. From the results in the inset of
Fig.~\ref{fig4} and in Fig.~\ref{fig6}, we can see that the fidelity decay
should become insignificant, with the populations of the qubit second-excited
state safely negligible.

The out-of-time-order correlator (OTOC)  is an important tool to study the
dynamics of quantum many-body systems. In Ref.~\cite{Braum2021}, a 3$\times$3
two-dimensional lattice of superconducting qubits is implemented to study the
time-reversibility, measurement of OTOC, propagation of quantum information, and
many-body localization in the presence of frequency disorder. The experiment is
found to be fairly described by the hard-core boson model in the time
range of $\sim$100 ns with $J/2\pi$ = 8.1 MHz and $U/2\pi$ = 244 MHz. The
initial states with one, two, and three excited particles in the 9-qubit lattice
are similar with the present $\psi_1$, $\psi_2$, and $\psi_3$ in one dimension.
In particular, the error of the OTOC (or equivalently the commutator) caused by
the qubit high level is carefully estimated considering the $ZZ$ interaction in
both cases of the 3$\times$3 lattice and a 7-qubit chain~\cite{Braum2021}. It is
found that the error would grow linearly with time, and in the cases of the
initial states with filled particle number ranging from two to five, the time
for reliable results is around a few hundreds to tens of ns for both the lattice
and chain. These estimations from the commutator, with somewhat different $U/J$
and the ratio of particle to qubit numbers, are comparable to the data in
Figs.~\ref{fig3} and \ref{fig6} from the fidelity calculations.

In a recent experiment with a superconducting 10-qubit chain, the time reversal
of the system is realized by using Floquet driving and the OTOCs are
successfully measured~\cite{Zhao21}. The sample parameters are listed in the
Supplemental Material~\cite{SM}. The experimental results show distinct
behaviours with and without a signature of information scrambling in the near
integrable system for the butterfly operators of $\sigma_x$ and $\sigma_z$,
respectively. For the measurements in the two cases, the initial states of
$\psi_4$ = $\ket{++++++++++}$ and $\psi_5$ = $\ket{0101010101}$ are prepared. It
is found that to well describe the experimental results, the consideration of
three qubit levels is necessary for $\psi_4$ while the two-level model modified
with $ZZ$ interaction can be used for $\psi_5$. Such difference is reasonable
from the data in Fig.~\ref{fig2} (also the inset of Fig.~\ref{fig4}) considering
that the fidelity decays faster for $\psi_4$ than for $\psi_5$ with the other
parameters unchanged. The failure of the simple two-level approximation is also
expected since the fidelity decreases almost to zero in the timescale of the
experiment (consideration of next nearest neighbour coupling will bring a small
change). As is discussed in section \ref{TR}, the fidelity decay with Floquet
driving is basically governed by the qubit actual coupling strength $J/2\pi$ =
10.8 MHz, which is much larger than the effective coupling strength $J/2\pi$ =
$\pm$3.8 MHz that determines the experimental timescale. In addition, the
fidelity in some cases shows a clear decrease while the population almost returns
to the initial values at the end of time evolution. This indicates a more
precise measure of fidelity than that of population~\cite{nie10,haa18}. In the
OTOC experiment with $\sigma_z$ butterfly operator and N\'{e}el initial
state~\cite{Zhao21}, the final measurement is the expectation value of
$\sigma_z$, which is equivalent to the qubit population.

Finally, we mention two experiments in the presence of transverse field with
superconducting circuits~\cite{Xu2020,Chen2021}.  In one experiment, dynamical
phase transitions in the two-level Lipkin-Meshkov-Glick model are observed with
16 all-to-all connected superconducting qubits~\cite{Xu2020}. The sample has a
small mean value of the coupling strength $J/2\pi$ = 1.45 MHz and an on-site
interaction $U/2\pi$ around 240 MHz, for which negligible fidelity decay is
expected. In another experiment, strong and weak thermalization is observed for
different initial states fully polarized or located on the equator of the Bloch
sphere, which is described by the two-level spin model considering the $Z$- and
$XY$-crosstalk induced disorders in chemical potential and local transverse
field~\cite{Chen2021}. The coupling strength $J/2\pi$ and on-site interaction
$U/2\pi$ of the sample are around  13 MHz and 235 MHz, respectively. Generally,
the fidelity decay for the initial state on the equator should be less affected
by the high-level participation with applied transverse field, as is shown in
Figs.~\ref{fig5} and \ref{fig7}. However, the comparison with the present
numerical simulation is not straightforward due to the two kinds of disorders
existed in the experiment.

\section{Summary and outlook}

Fidelity is a standard and precise measure of the distance between quantum
states~\cite{nie10,haa18}, which was used to investigate the accuracy and
validity of the two-level approximation for the multilevel transmons in the QQD
experiments. In order to evaluate the effect of the high energy levels of
transmon qubits, the fidelity decay (i.e., the time dependent overlap of wave
functions) was calculated for two experimental situations with and without time
reversal process and for a number of system Hamiltonians and initial states. Our
results showed that the qubit high-level populations may lead to remarkable
fidelity decay during state evolution in a QQD experiment, depending on the
ratio $U/J$, the initial states, and the applied driving. In many cases, it is
seen that the ratio of $U/J$ should roughly be larger than $\sim$ 60 for safely
neglecting the high-level populations created in the quench process in a
timescale of 2$\pi/J$. For the usual transmon qubits with $U/2\pi$ $\sim$ 240
MHz, for instance, the qubit coupling strength $J/2\pi$ should be $\sim$ 4 MHz
and below for a simulation experiment with time evolution up to $\sim$ 250 ns.
The criterion is nevertheless relative and under these conditions, the fidelity
generally remained above 0.9 and the high-level population for individual qubits
was well below 0.01 in most cases. Furthermore, our results showed that the
fidelity decay also depends on the initial state and system Hamiltonians applied
with external drivings. These factors need to be taken into account for the
application of the two-level spin and hard-core boson models in the QQD
experiments.

Various many-body quantum phenomena have been studied and successfully
demonstrated in a number of QQD experiments using transmon qubits with
sufficiently low high-level populations~\cite{Xu2018, Yan2019, Xu2020, Gong2021,
	Chen2021, Braum2021, Kara2022}. The key parameter for the applicability of the
two-level approximation for transmons is the ratio $U/J$ since the ideal
hard-core boson model is achieved when $U/J$ tends to infinity. From the above
discussions, satisfactory approximations can be made experimentally taking
advantage of the present-day technologies of device design and fabrication. In
particular, we mention the strong anharmonicity or high $U$ devices such as
capacitively shunted qubits, which have become available~\cite{Yan16,Yur21} and
may offer a convenient playground in the future for the study of the spin and
hard-core boson models using superconducting circuits.


\section*{Acknowledgments}

This work was partly supported by the Key-Area Research and Development Program
of GuangDong Province (Grant No. 2018B030326001), and the National Natural
Science Foundation of China (Grant No.~11874063).  H. F. Y. acknowledges supports
from the NSF of Beijing (Grant No.~Z190012) and the National Natural Science
Foundation of China (Grant No.~11890704). H. F. acknowledges supports from the
National Natural Science Foundation of China (Grant Nos.~11934018 and T2121001),
Strategic Priority Research Program of Chinese Academy of Sciences (Grant
No.~XDB28000000), and Beijing Natural Science Foundation (Grant No.~Z200009).

\clearpage \widetext
\begin{center}
	{\bf Supplemental Material for \\ ``Two-level approximation of transmons in quantum quench experiments"}
\end{center}

\setcounter{equation}{0} \setcounter{figure}{0}
\setcounter{table}{0} \setcounter{page}{1} \setcounter{secnumdepth}{3} \makeatletter
\renewcommand{\theequation}{S\arabic{equation}}
\renewcommand{\thetable}{S\arabic{table}}
\renewcommand{\thefigure}{S\arabic{figure}}
\renewcommand{\bibnumfmt}[1]{[S#1]}
\renewcommand{\citenumfont}[1]{S#1}

\makeatletter
\def\@hangfrom@section#1#2#3{\@hangfrom{#1#2#3}}
\makeatother


\maketitle

~~~

This Supplemental Material includes one table and twelve figures,  which are
explained below.


\begin{center}
{\bf I. Consideration of model parameters for numerical simulations}
\end{center}
	
In Table~\ref{table1}, we list the relevant device parameters of a ten transmon
qubit chain from Ref.~\cite{SZhao21} for the consideration of our model
parameters for numerical simulations. These parameters are typical and often
seen in the QQD experiments using superconducting
qubits~\cite{SYan2019,SBraum2021}. For transmon qubits, the coherence times are
much longer than the time scales of interest in this work, so dissipations are
not taken into account in our simulation. We note that the anharmonicity
$-U_j/2\pi$ given in the table is around $-212$ and $-264$ MHz for odd and even
number qubits, respectively (a negative sign is given explicitly for the
anharmonicity so we have a positive $U_j$ for the convenience of discussion).
These anharmonicity parameters are used for all calculations in this work except
for the results of Fig.~8 discussed in Sec.~V A of the main text. Generally, a
more smooth or constant $U_j/2\pi$ tends to slightly increase the fidelity decay
rate. On the other hand, an averaged $uniform$ nearest-neighbour (NN) coupling
strength $J/2\pi$ = 10.8 MHz is used for simplicity in the simulation of Floquet
driven system. In this case, the driving parameters are such that the forward
and backward state evolutions have the effective coupling strengths of 3.8 and
-3.8 MHz, respectively. Uniform NN couplings with varying strength are also used
for the simulations in other situations.

\begin{center}
	{\bf II. Detailed population variations of individual qubits}
\end{center}

\begin{figure*}[t]
	\includegraphics[width=1.25\textwidth,angle=-90]{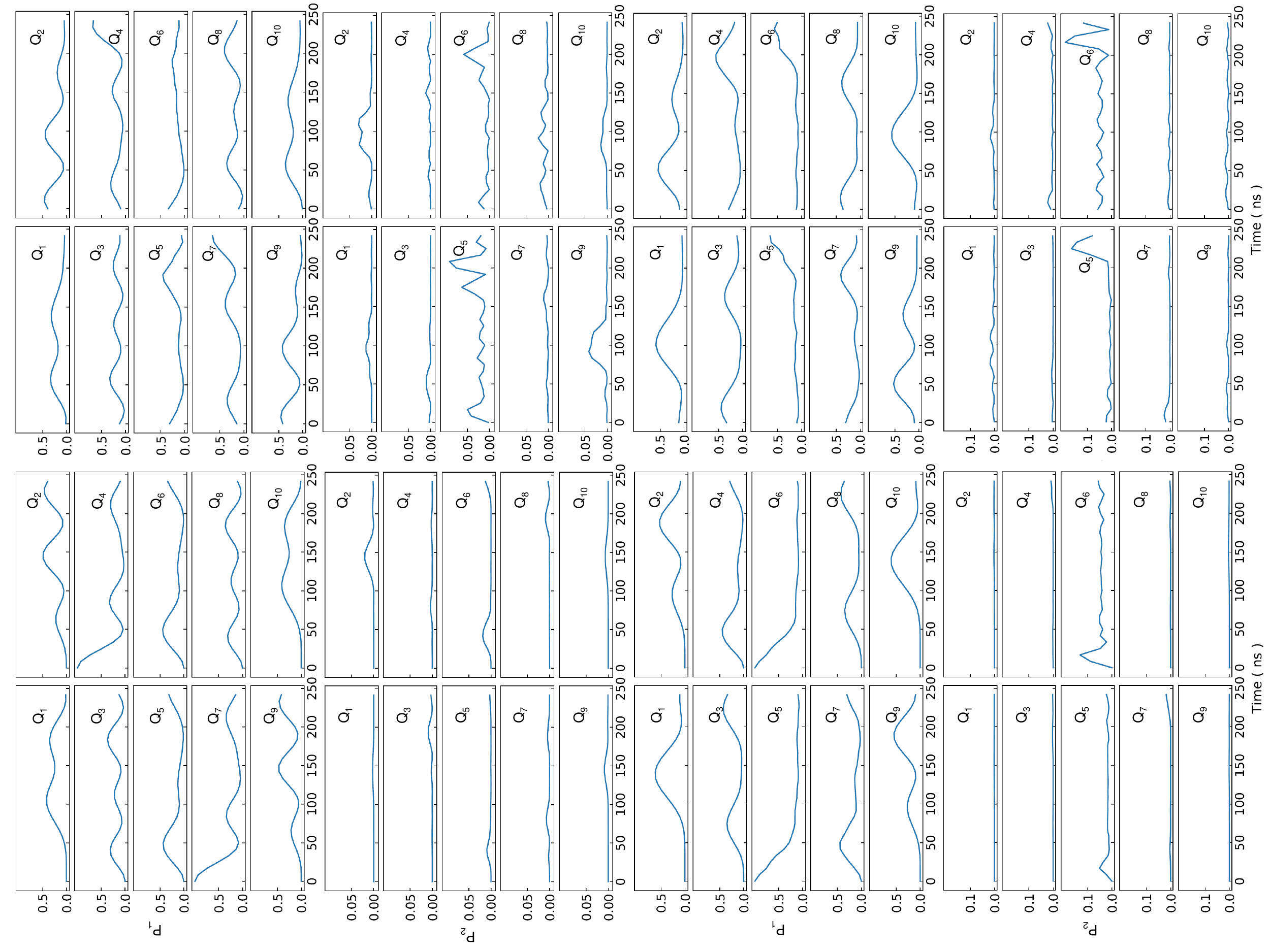} 
	\caption{Detailed population variations of individual qubits for the results in
		Fig.~2 with Floquet driving and initial states of $\ket{0001001000}$ (upper
		four panels) and $\ket{0000110000}$ (lower four panels). The left and right
		columns are the forward and backward evolutions, while the odd and even rows
		are for the first- and second-excited states, respectively.}
	\label{figS1a} 
\end{figure*}

\begin{figure*}[t]
	\includegraphics[width=1.25\textwidth,angle=-90]{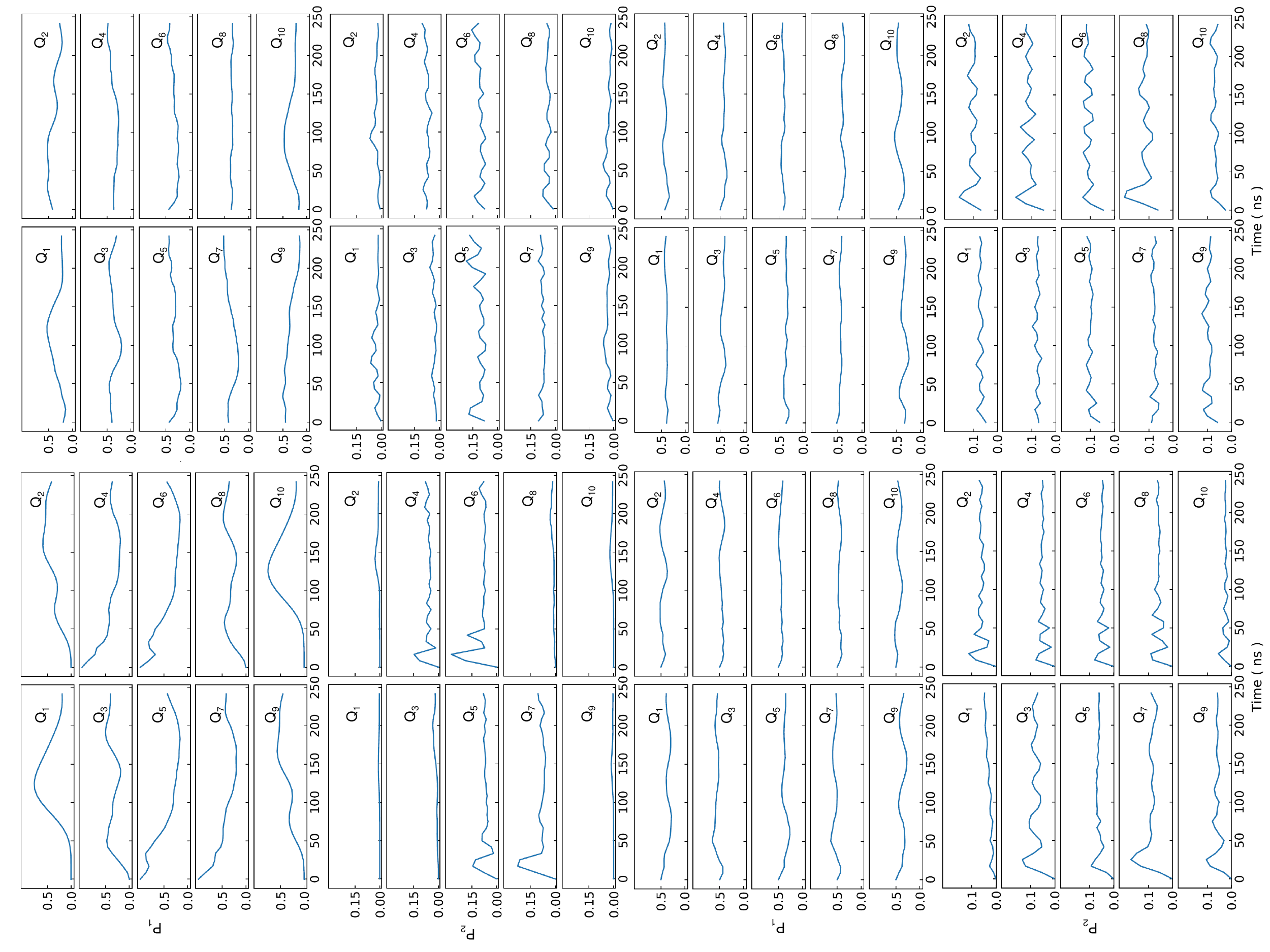} 
	\caption{Detailed population variations of individual qubits for the results in
		Fig.~2 with Floquet driving and initial states of $\ket{0001111000}$ (upper
		four panels) and $\ket{++++++++++}$ (lower four panels). The left and right
		columns are the forward and backward evolutions, while the odd and even rows
		are for the first- and second-excited states, respectively.}
	\label{figS1b} 
\end{figure*}

\begin{figure*}[t]
	\includegraphics[width=1.25\textwidth,angle=-90]{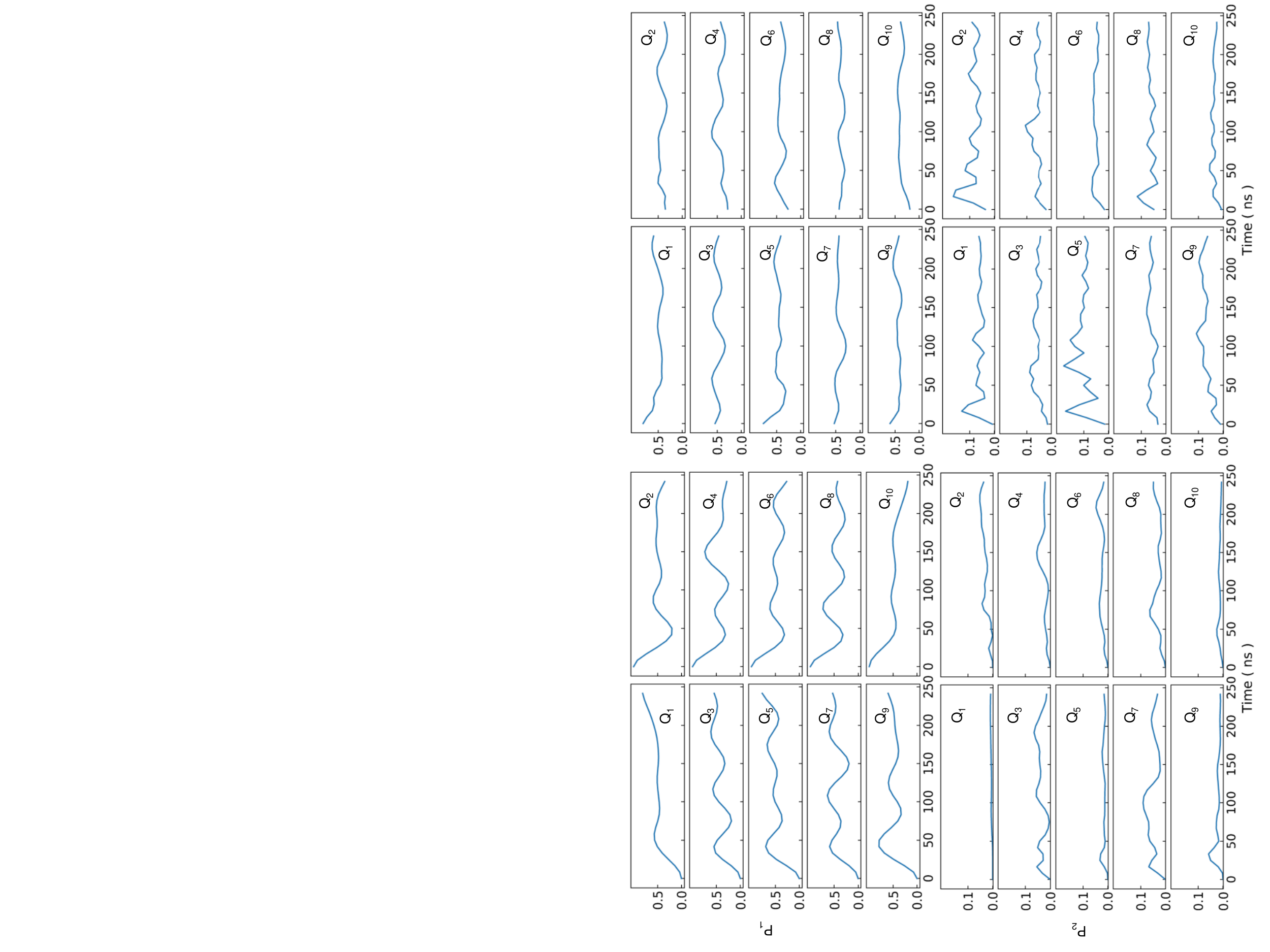} 
	\caption{Detailed population variations of individual qubits for the results in
		Fig.~2 with Floquet driving and initial states of $\ket{0101010101}$. The left
		and right columns are the forward and backward evolutions, while the upper and
		lower panels are for the first- and second-excited states, respectively.}
	\label{figS1c} 
\end{figure*}

\begin{figure*}[t]
	\includegraphics[width=1.25\textwidth,angle=-90]{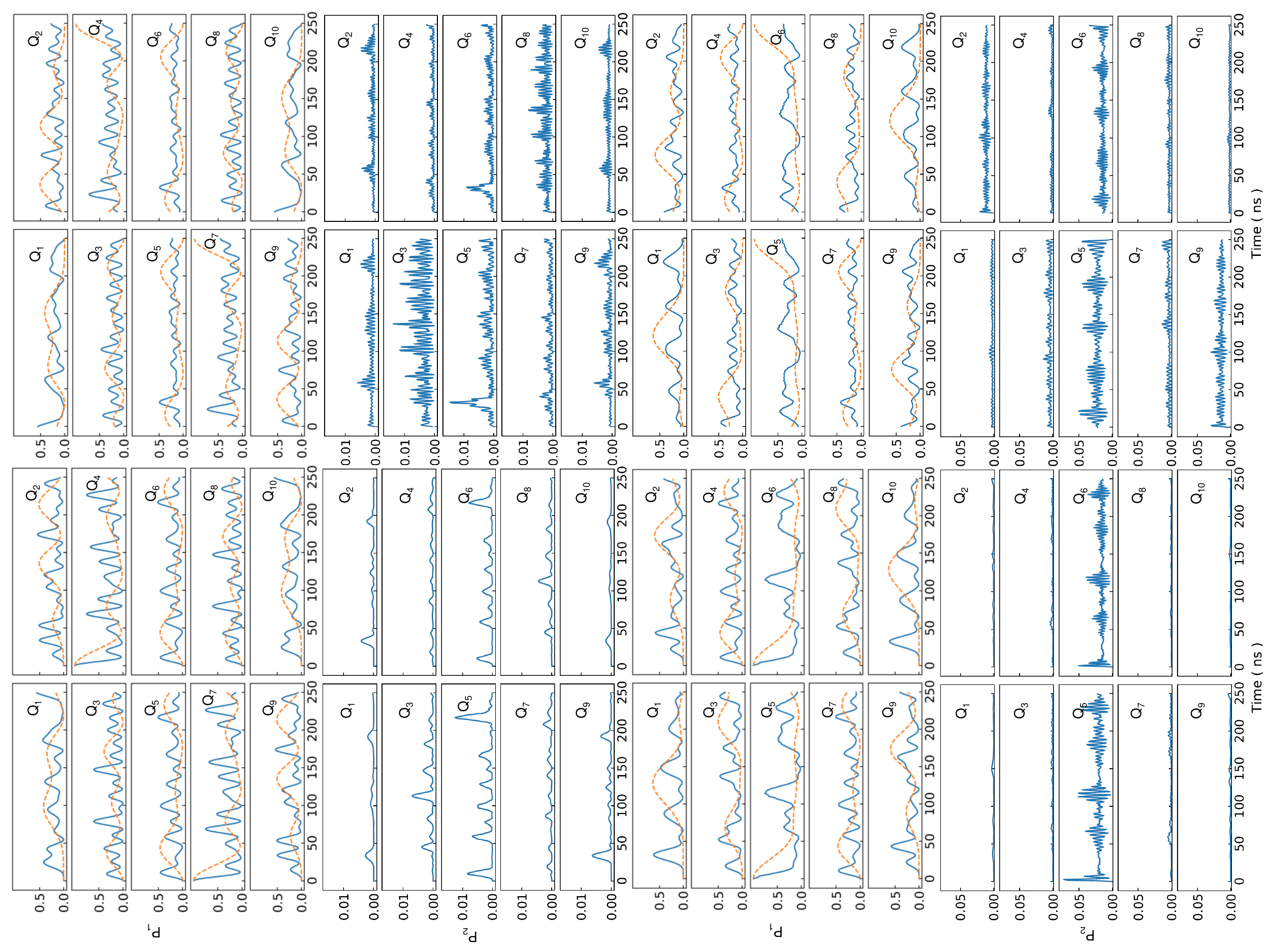} 
	\caption{Detailed population variations of individual qubits for the results in
		Fig.~3 (solid lines) with NN coupling of 16 MHz and initial states of
		$\ket{0001001000}$ (upper four panels) and $\ket{0000110000}$ (lower four
		panels). The left and right columns are the forward and backward evolutions,
		while the odd and even rows are for the first- and second-excited states,
		respectively. Dashed lines correspond to those in the inset of Fig.~3 with NN 
		coupling of 4 MHz.}
	\label{figS2a} 
\end{figure*}

\begin{figure*}[t]
	\includegraphics[width=1.25\textwidth,angle=-90]{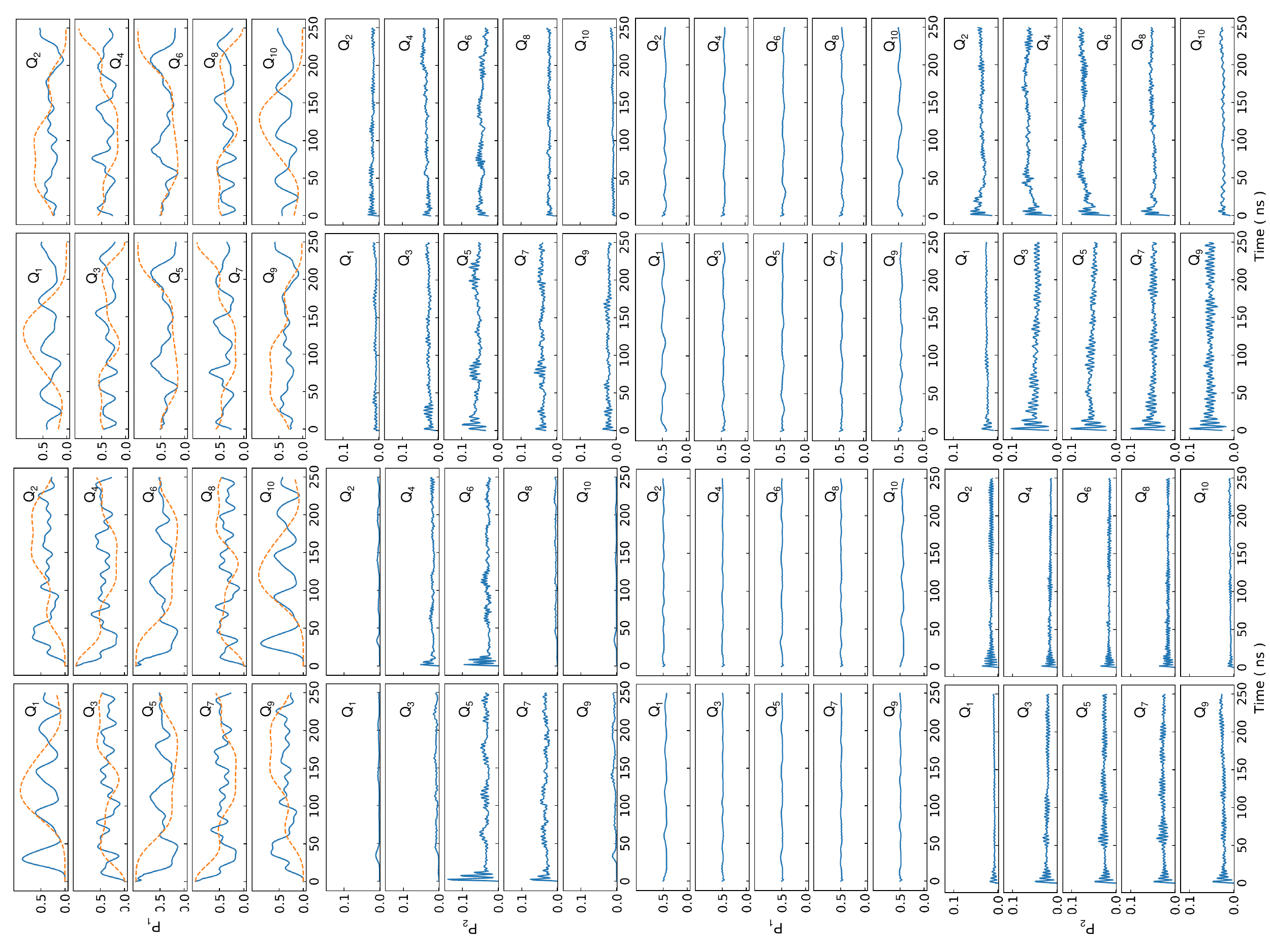} 
	\caption{Detailed population variations of individual qubits for the results in
		Fig.~3 (solid lines) with NN coupling of 16 MHz and initial states of
		$\ket{0001111000}$ (upper four panels) and $\ket{++++++++++}$ (lower four
		panels). The left and right columns are the forward and backward evolutions,
		while the odd and even rows are for the first- and second-excited states,
		respectively. Dashed lines correspond to those in the inset of Fig.~3 with 
		NN coupling of 4 MHz.}
	\label{figS2b} 
\end{figure*}

\begin{figure*}[t]
	\includegraphics[width=1.25\textwidth,angle=-90]{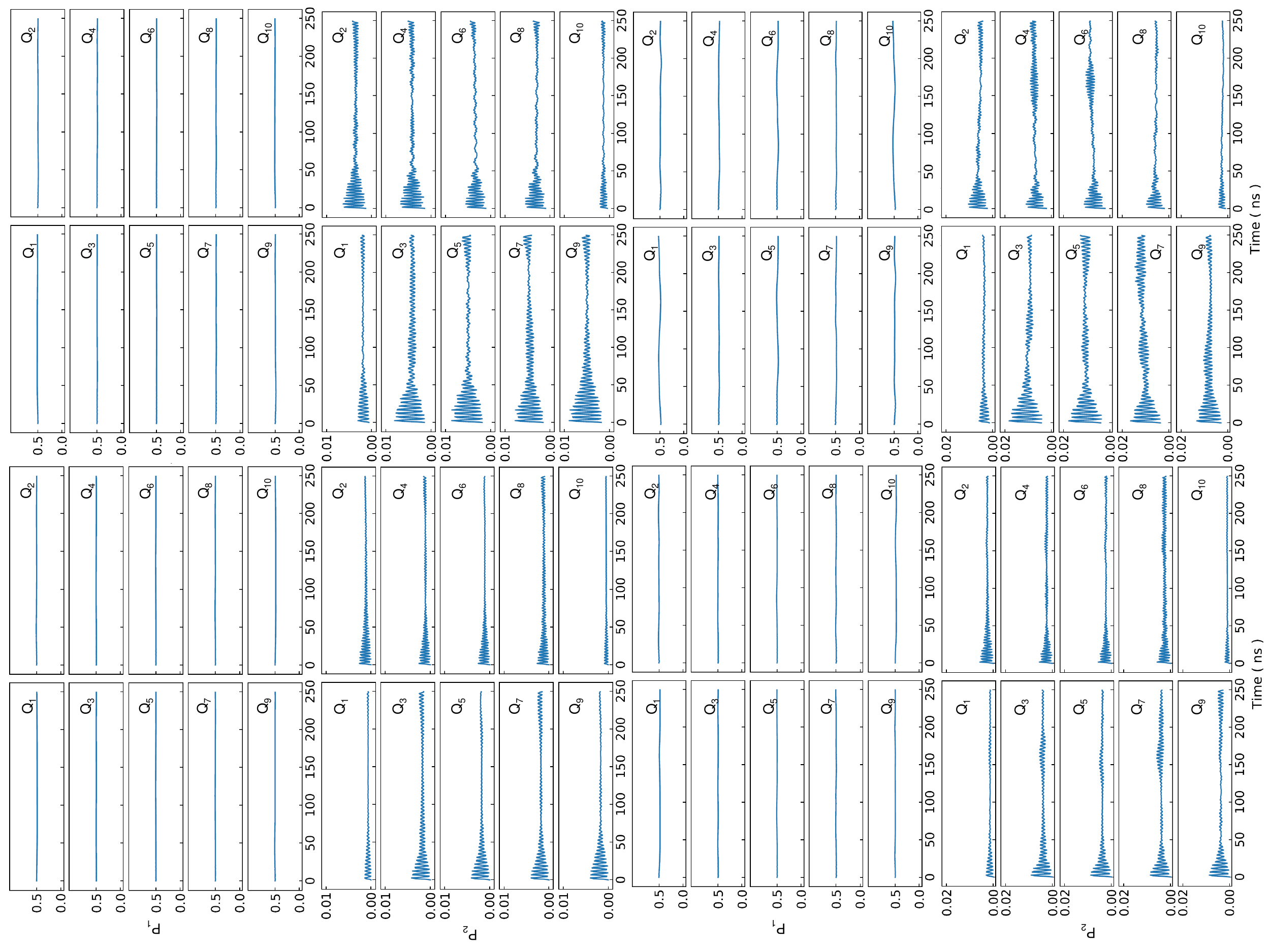} 
	\caption{Detailed population variations of individual qubits for the results in
		Fig.~4 with initial state $\ket{++++++++++}$ and NN coupling of 4 MHz (upper
		four panels) and 6 MHz (lower four panels). The left and right columns are the
		forward and backward evolutions, while the odd and even rows are for the first- and
		second-excited states, respectively.}
	\label{figS3a} 
\end{figure*}

\begin{figure*}[t]
	\includegraphics[width=1.25\textwidth,angle=-90]{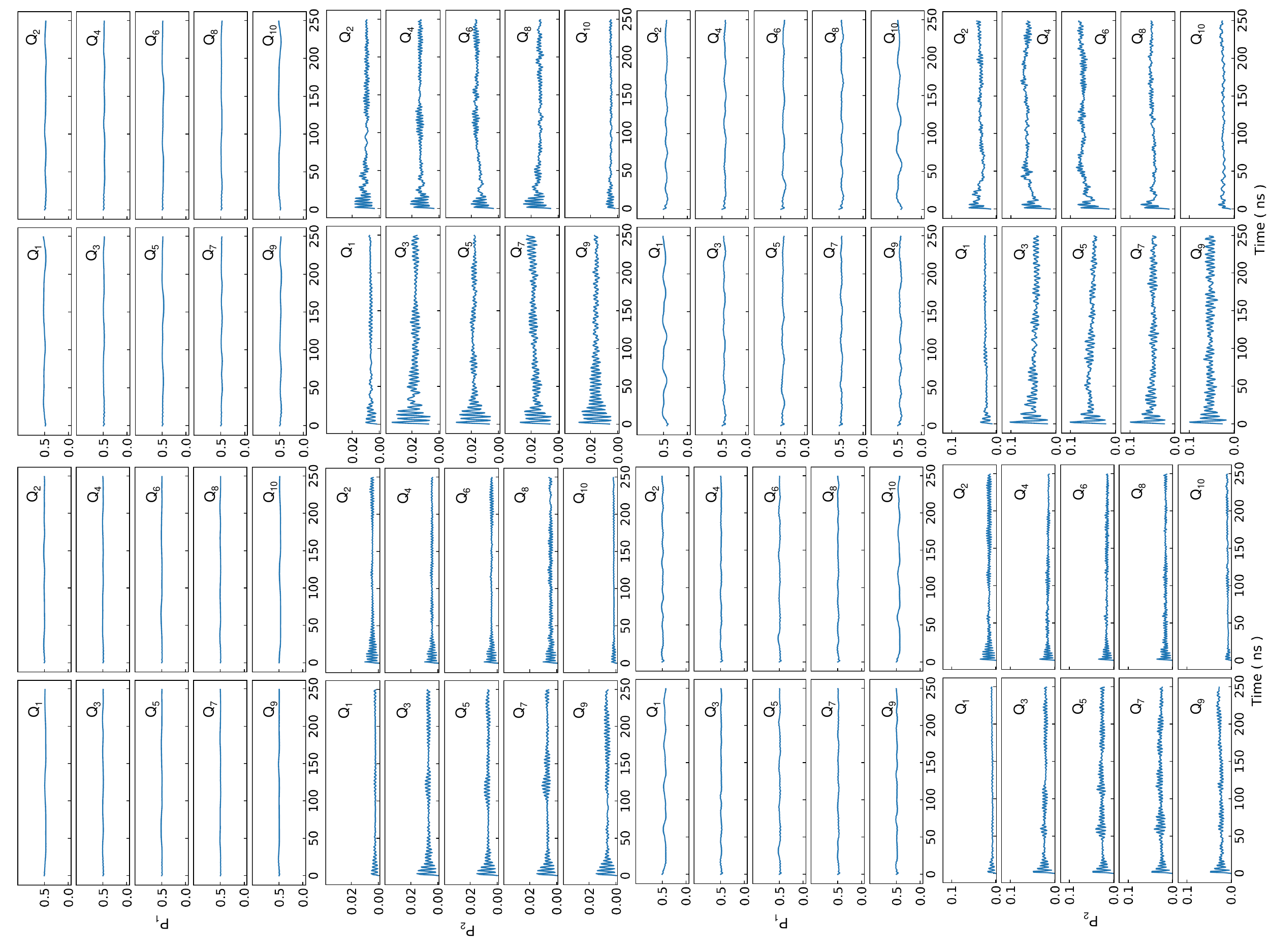} 
	\caption{Detailed population variations of individual qubits for the results in
		Fig.~4 with initial state $\ket{++++++++++}$ and NN coupling of 8 MHz (upper
		four panels) and 16 MHz (lower four panels). The left and right columns are the
		forward and backward evolutions, while the odd and even rows are for the first- 
		and second-excited states, respectively.}
	\label{figS3b} 
\end{figure*}

\begin{figure*}[t]
	\includegraphics[width=1.25\textwidth,angle=-90]{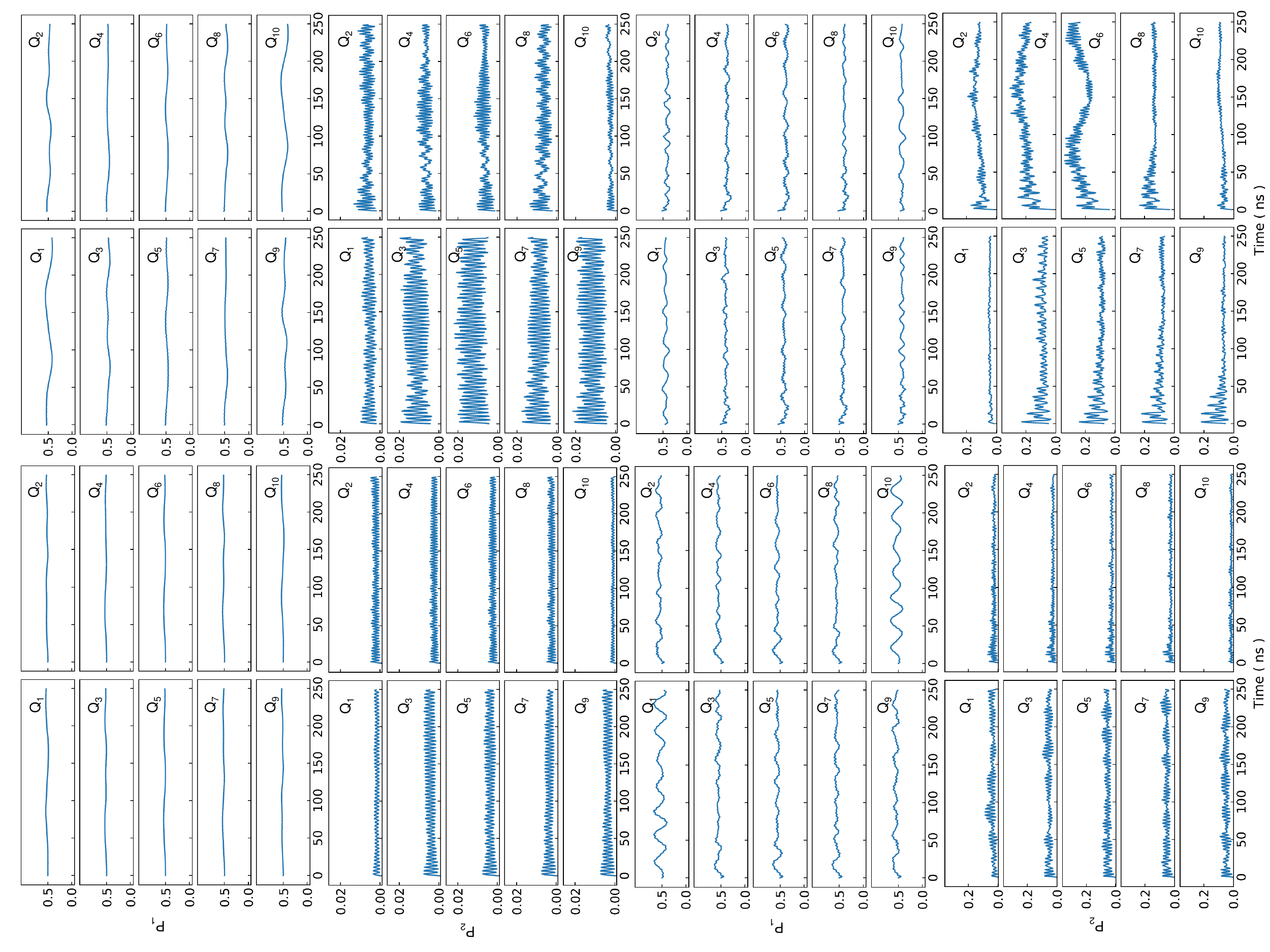} 
	\caption{Detailed population variations of individual qubits for the results in
		Fig.~5 with initial state $\ket{++++++++++}$. The NN coupling and transverse
		field strength are 4 and 4 MHz in upper four panels, and 16 and 16 MHz in lower
		four panels. The left and right columns are the forward and backward
		evolutions, while the odd and even rows are for the first- and second-excited
		states, respectively.}
	\label{figS4a} 
\end{figure*}

\begin{figure*}[t]
	\includegraphics[width=1.25\textwidth,angle=-90]{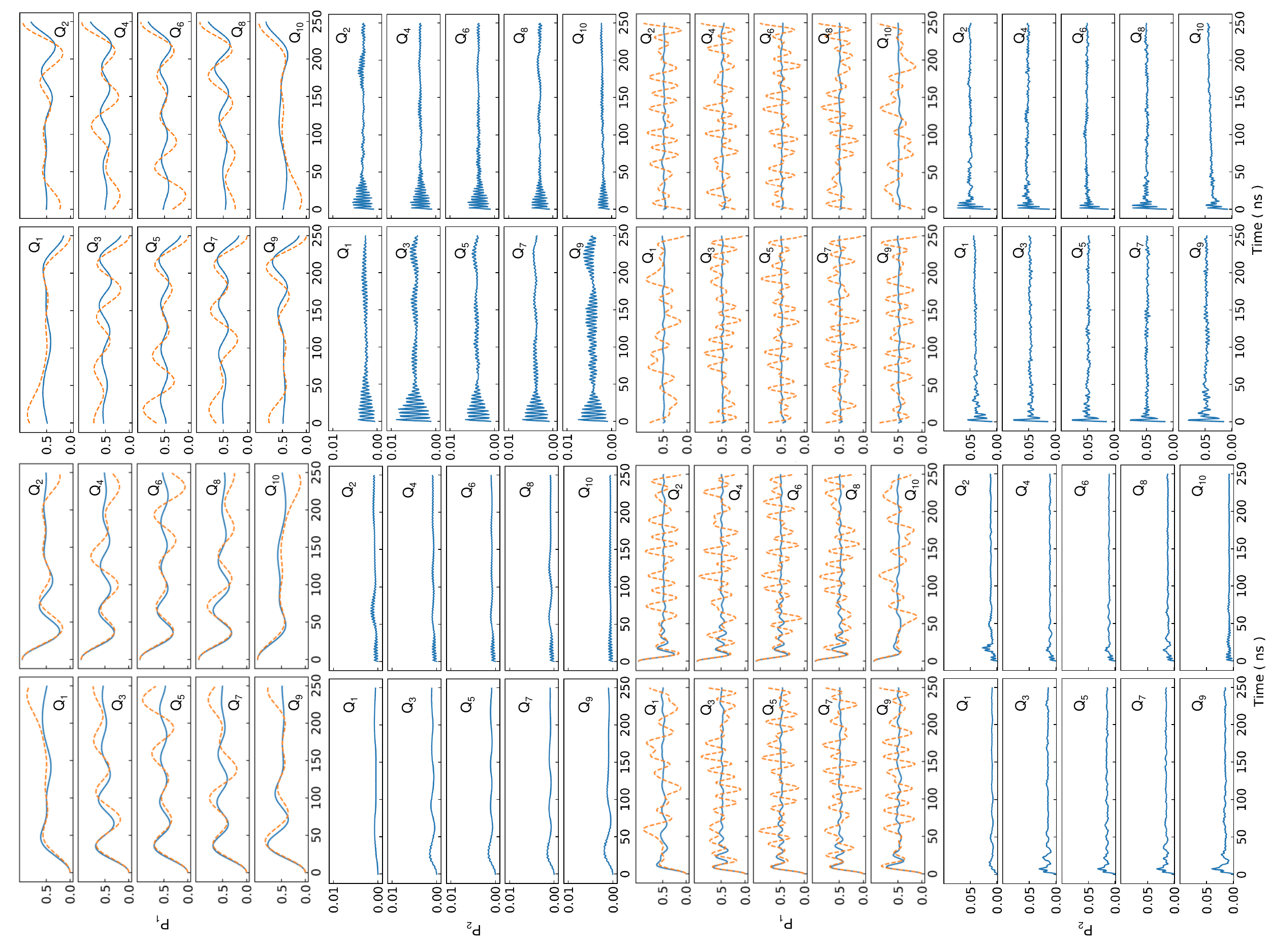} 
	\caption{Detailed population variations of individual qubits for the results in
		Fig.~5 with initial state $\ket{0101010101}$. The NN coupling and transverse
		field strength are 4 and 4 MHz in upper four panels, and 16 and 16 MHz in lower
		four panels. The left and right columns are the forward and backward
		evolutions, while the odd and even rows are for the first- and second-excited
		states, respectively. The dashed lines are the corresponding results without
		transverse field.}
	\label{figS4b} 
\end{figure*}

\begin{figure*}[t]
	\includegraphics[width=0.99\textwidth]{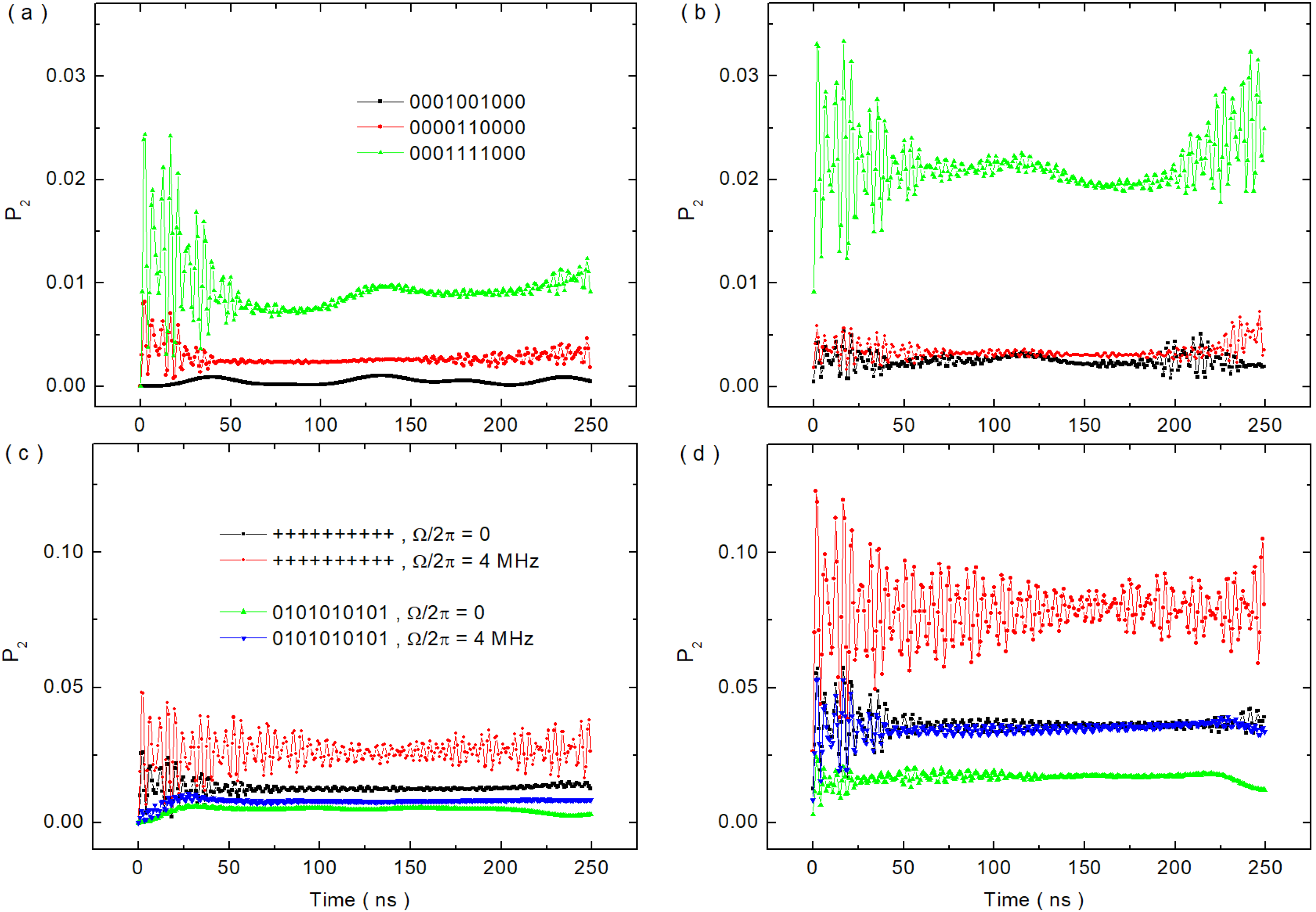} 
	\caption{Time dependence of total populations of the second-excited state with
		$J/2\pi$ = 4 MHz for (a, b) $\psi_1$, $\psi_2$, and $\psi_3$, (c, d) $\psi_4$
		and $\psi_5$. In (a, c) and (b, d) are the forward and backward evolutions,
		respectively.}
	\label{figS5} 
\end{figure*}

\begin{figure*}[t]
	\includegraphics[width=0.99\textwidth]{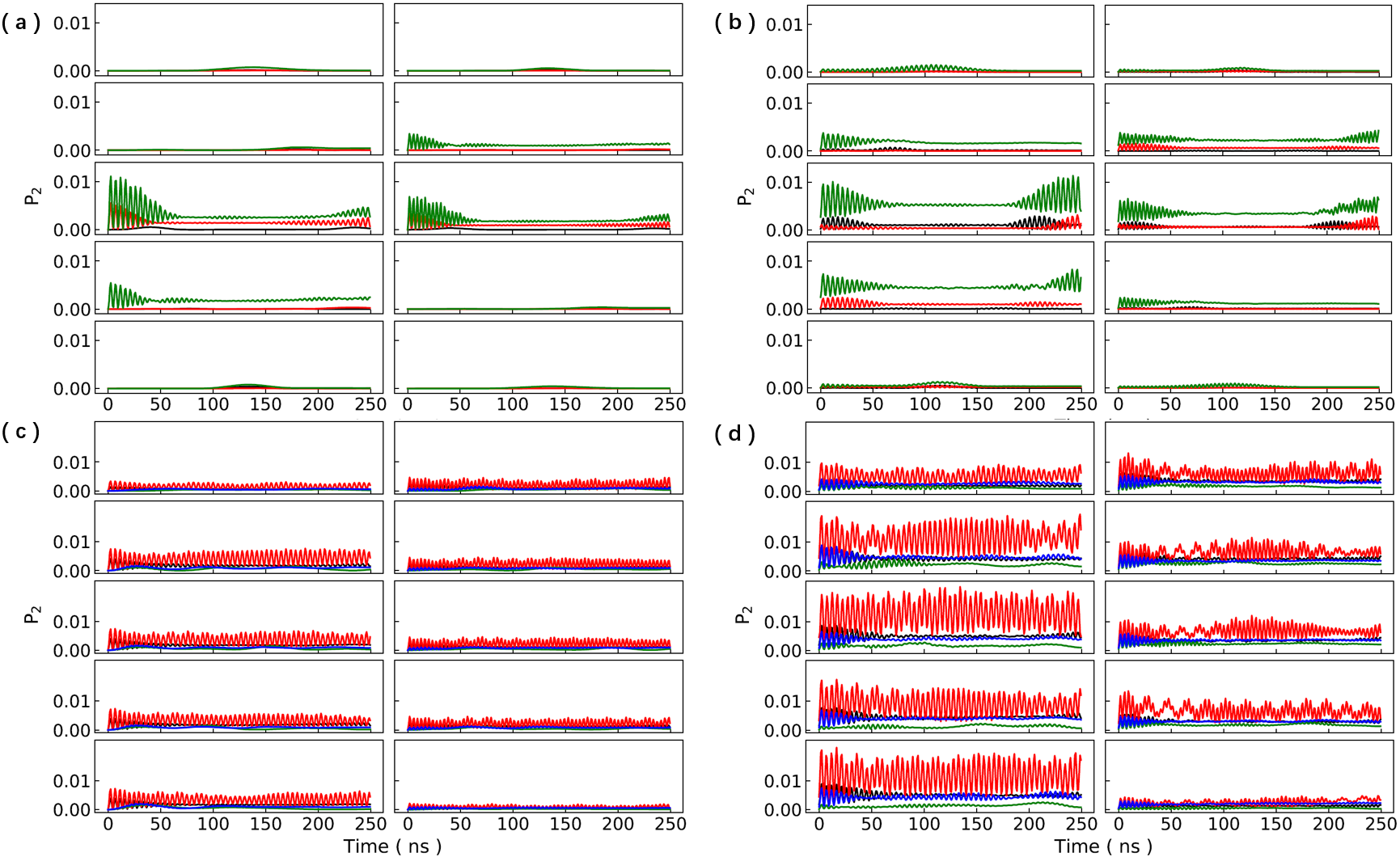} 
	\caption{Detailed population variations of the second-excited state for 
		individual qubits with $J/2\pi$ = 4 MHz for (a, b) $\psi_1$, $\psi_2$, and
		$\psi_3$, (c, d) $\psi_4$ and $\psi_5$. In (a, c) and (b, d) are the forward
		and backward evolutions, respectively. All curves have a one-to-one
		correspondence to those in Fig.~\ref{figS5} with the same colour.}
	\label{figS6} 
\end{figure*}

In Fig.~\ref{figS1a} through Fig.~\ref{figS4b}, we show the detailed time
dependence of populations for individual qubits during both the forward and 
backward evolutions. Figs.~\ref{figS1a} to \ref{figS1c}, Figs.~\ref{figS2a} and
\ref{figS2b}, Figs.~\ref{figS3a} and \ref{figS3b}, and Figs.~\ref{figS4a} and
\ref{figS4b} are the data corresponding to the cases in Fig.~2, Fig.~3 (inset),
Fig.~4, and Fig.~5 in the main text, respectively. The detailed population
variations with time of the second-excited state for weak coupling strength
$J/2\pi$ = 4 MHz are shown separately in Fig.~\ref{figS5} and Fig.~\ref{figS6}
due to their much smaller values.

\begin{center}
	{\bf III. Entanglement entropy}
\end{center}

\begin{figure*}[t]
	\includegraphics[width=0.99\textwidth]{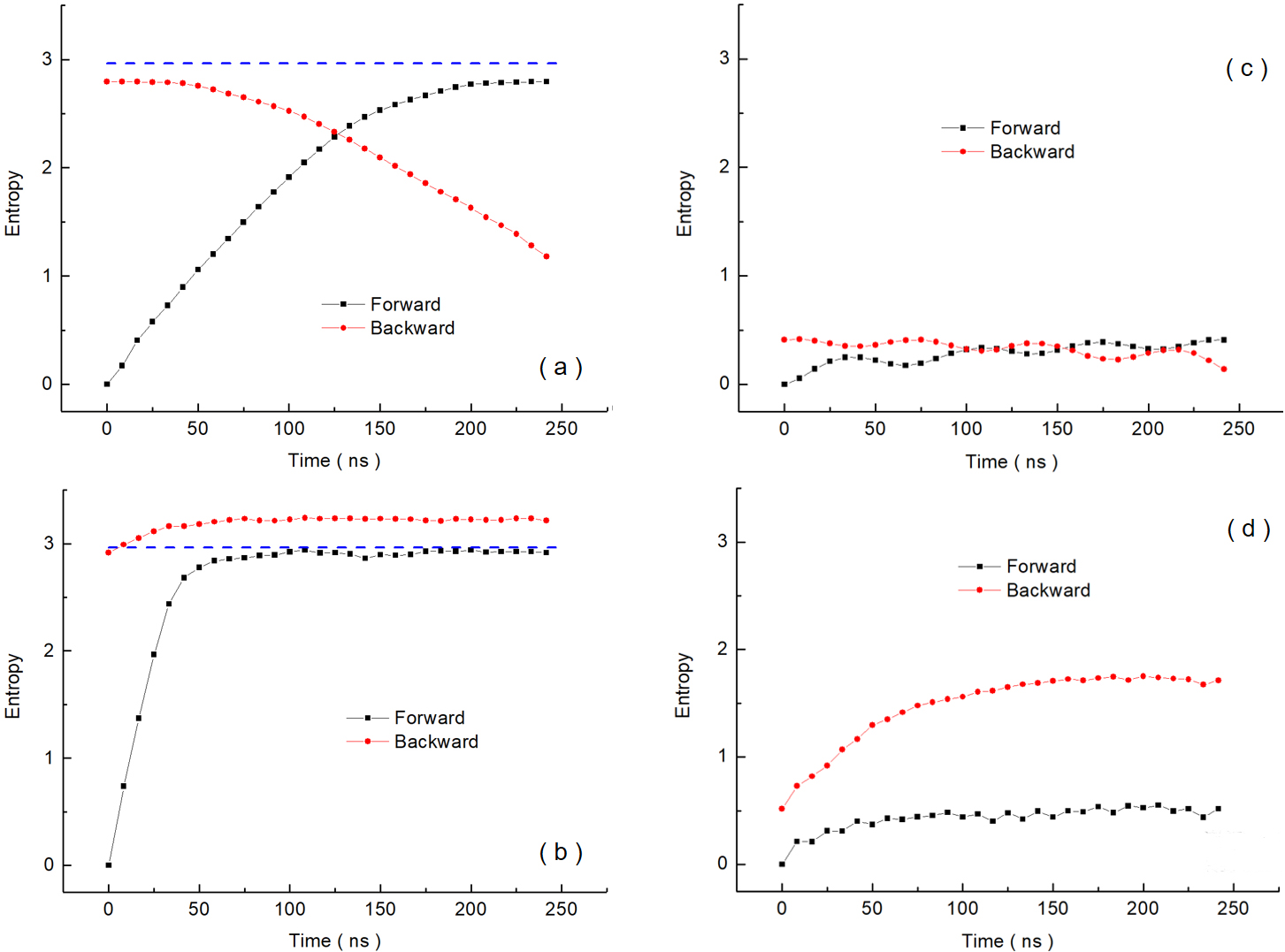} 
	\caption{Entanglement entropies (of half-chain subsystem) corresponding to the
		data in Fig.~5 (see also Figs.~\ref{figS4a} and \ref{figS4b}), with initial
		states $\ket{0101010101}$ in (a) and (b), and $\ket{++++++++++}$ in (c) and
		(d). The parameters $J/2\pi$ and $\Omega/2\pi$ are 4 MHz for (a) and (c), and
		16 MHz for (b) and (d). Dashed lines in (a) and (b) are the Page value
		$S^{Page}$ = [10ln(2)-1]/2 for the 10-qubit chain.}
	\label{figS7} 
\end{figure*}

The entanglement entropies corresponding to the data in Fig.~5 in the main text
are shown in  Fig.~\ref{figS7}.

\newpage

\begin{table*}[h]
	\caption{Typical transmon qubit device parameters~\cite{SZhao21} considered for
		numerical simulations in this work, where $-U$ is the qubit anharmonicity and
		$J_{j,j+1}$ is the coupling strength between the nearest-neighbour (NN)
		qubits Q$_j$ and Q$_{j+1}$.}
	\label{table1} 
	\begin{ruledtabular}
		\setlength{\tabcolsep}{9pt}
		\centering
		\resizebox{\textwidth}{!}{
			\begin{tabular}{c|cccccccccc}
				& Q$_1$ & Q$_2$ & Q$_3$ & Q$_4$ & Q$_5$ & Q$_6$ & Q$_7$ & Q$_8$ & Q$_9$ & Q$_{10}$ \\
				\hline
				$-U/2\pi$ (MHz)  & -212 & -264 & -210 & -268 & -212 & -268 & -214 & -264 & -214 & -264 \\
				$J_{j,j+1}/2\pi$ (MHz) & \multicolumn{10}{c} {  \hspace{-6mm} 10.72 \hspace{6mm} 10.73 \hspace{6mm} 10.99 \hspace{6mm} 11.05 
					\hspace{6mm} 10.88 \hspace{6mm} 10.48 \hspace{6mm} 10.86 \hspace{6mm} 10.79 \hspace{6mm} 10.78}\\%
		\end{tabular}}
	\end{ruledtabular}
\end{table*}


\begin{thebibliography}{99}
\bibitem{rem1}Further details will be presented in a separate work.
\end{thebibliography}
\end{document}